\documentclass{ws-ijmpa}
\usepackage{slashed}
\usepackage{axodraw}
\usepackage{wasysym}

\newcommand{\ltsimeq}{\raisebox{-0.6ex}{$\ \stackrel
           {\raisebox{-.2ex}{$\textstyle <$}}{\sim}\ $}}
\newcommand{\gtsimeq}{\raisebox{-0.6ex}{$\ \stackrel
           {\raisebox{-.2ex}{$\textstyle >$}}{\sim}\ $}}

\begin{document}

%\bibliographystyle{unsrt}

%\preprint{hep-ph/0609055}

\markboth{Douglas M. Gingrich}
{Signatures of Singlet Neutrinos in Large Extra Dimensions at the LHC}

\catchline{}{}{}{}{}

\title{SIGNATURES OF SINGLET NEUTRINOS IN LARGE EXTRA DIMENSIONS AT THE LHC}

\author{DOUGLAS M. GINGRICH\footnote{Also at TRIUMF, Vancouver, BC V6T
2A3 Canada}} 

\address{Centre for Particle Physics, Department of Physics,\\
University of Alberta, Edmonton, AB T6G 2G7 Canada\\
gingrich@ualberta.ca}

\maketitle

\begin{history}
%\received{\today}
\received{10 July 2009}
\revised{27 August 2009}
\end{history}

%%%%%%%%%%%%%%%%%%%%%%%%%%%%%%%%%%%%%%%%%%%%%%%%%%%%%%%%%%%%%%%%%%%%%%%%%%%%%%%
\begin{abstract} 
It is a challenge to explain why neutrinos are so light compared to
other leptons.  
Small neutrino masses can be explained if right-handed fermions
propagate in large extra dimensions.  
Fermions propagating in the bulk would have implications on Higgs boson
decays. 
If the Higgs boson is discovered at the Large Hadron Collider (LHC), a
detailed analysis may reveal the presence of large extra dimensions.
This paper reviews the status of large extra-dimensional models in the
context of the current limits on Higgs boson masses and the fundamental
Planck scale in extra dimensions.  

\end{abstract}

%\ccode{PACS numbers: 04.70.Bw, 04.50.+h, 12.60.-i, 04.70.-s}

%%%%%%%%%%%%%%%%%%%%%%%%%%%%%%%%%%%%%%%%%%%%%%%%%%%%%%%%%%%%%%%%%%%%%%%%%%%%%%%
\section{Introduction} \label{sec1}

Neutrino oscillation experiments suggest that neutrinos have a very
small but nonzero mass.\cite{Kam98}$^-$\cite{SNO05}
To explain why neutrinos are so light compared to other leptons is a
challenge.
The traditional approach is to give neutrinos small mass via a seesaw
mechanism. 
In the type-I seesaw approach, a large right-handed Majorana mass $M_R$
suppresses one of the eigenvalues of the neutrino mass matrix.
This leads to a neutrino mass $m_\nu \sim m_\mathrm{D}^2 / M_R$, where 
$m_\mathrm{D}$ is the mass of a light Dirac fermion.
The neutrino mixing required to explain the atmospheric, solar, and
accelerator neutrino oscillation data requires a super-heavy energy
scale for $M_R$. 
Explaining the neutrino masses and oscillations is one of the grand
challenges of particle physics.

Another interesting challenge is to explain the hierarchy problem: the
fine tuning required to maintain a low electroweak scale in the presence
of the Planck scale. 
Supersymmetry, technicolour, and extra dimensions have all been used to
address the hierarchy problem. 
In particular, the extra-dimensions paradigm can lead to the possibility
of low-scale gravity.\cite{Arkani98a}$^-$\cite{Arkani99}    
In the large extra-dimensions approach (often referred to as ADD), the
Standard Model (SM) fields are usually localised on (3+1)-dimensional
wall (a 3-brane), while gravity is allowed to propagate in all of the
dimensions (the bulk).  
If the fundamental gravity scale is about a TeV, the ultraviolet cut-off
for quantum corrections to the Higgs boson mass is also about a TeV.
This recasts the hierarchy problem in terms of geometry and the
stabilisation of large extra dimensions.

In the large extra-dimensions paradigm, small neutrino masses can be
generated without implementing a super-heavy energy scale, as done in
a seesaw mechanism\footnote{Higher-dimensional seesaw mechanisms are
also possible.\cite{Dienes99}}.   
Small neutrino masses are naturally explained if at least one
right-handed fermion propagates in the extra
dimensions.\cite{Dienes99}$^-$\cite{Arkani01} 
Since the SM gauge fields are localised to the 3-brane, a bulk fermion
must be a SM singlet: ``bulk right-handed neutrino'' or ``singlet
right-handed neutrino''. 
Right-handed neutrinos can freely propagate in the extra dimensions
because they have no quantum numbers to constrain them to the SM 3-brane. 
Therefore, they can also be classified as ``sterile'' neutrinos. 
The bulk neutrinos couple to the brane-localised SM fields with small
Yukawa couplings.  
The couplings are small because of the large relative volume of the
bulk manifold compared to the thin SM 3-brane.
However, because of mixing with a large number of Kaluza Klein (KK)
states in the bulk, the interaction probability with the SM fields can
be enhanced. 
Thus the effect of bulk neutrinos on, for example,  Higgs boson decays
can be significant.   

This paper is organised as follows.
Section~2 considers some of the experimental results from neutrino
experiments that we will use. 
Section~3 reviews the ideas of right-handed neutrinos in large extra
dimensions.  
The neutrino mass and coupling are discussed in subsection~3.1 and
subsection~3.2, respectively.
Subsection~3.3 presents bulk neutrinos in other spaces and the effect of
compactification.
Constraints on the size and number of extra dimensions are discussed
in subsection~3.4.
Section~4 identifies some of the signatures of right-handed neutrinos in
large extra dimensions, which could be revealed by experiments at the
LHC.  
First examined are the implications for the $\tau$ lepton decay of a
heavy charged Higgs boson in subsection~4.1.
Then, the invisible decay mode of a light Higgs boson is explored in
subsection~4.2.   
Finally, section~5 summarises our findings.
With the startup of the LHC, such detailed phenomenological reviews are
timely and of value. 

%%%%%%%%%%%%%%%%%%%%%%%%%%%%%%%%%%%%%%%%%%%%%%%%%%%%%%%%%%%%%%%%%%%%%%%%%%%%%%%
\section{Some Neutrino Oscillation and Mass Results} \label{sec2}

Nearly all the useful measurements of neutrino properties, that we will
be concerned with, come from neutrino flavour oscillation experiments.
Neutrino oscillations imply that the neutrinos have mass and that their
lepton flavour is mixed.
From neutrino oscillation experiments, one can deduce the absolute
difference in the square of neutrino masses.  
It is common to associate measurements from solar neutrino oscillation
experiments with $\Delta m_{\astrosun}^2 = \Delta_{21}^2 = m_2^2
-m_1^2$ and $\theta_{\astrosun} = \theta_{12}$, and measurements from
atmospheric neutrino experiments with $\Delta m_\mathrm{atm}^2 =
\Delta_{32}^2 = |m_3^2 -m_2^2|$ and $\theta_\mathrm{atm} = \theta_{23}$,
where $m_1, m_2,$ and $m_3$ are the neutrino physical masses.  
The LSND results\cite{LSND} are not considered in this paper.
These results have been shown to be incompatible with solar and
atmospheric oscillation data in models of bulk
neutrinos.\cite{Mohapatra00}  

The experimental oscillation measurements lead to the following results,
which are summarised by the Particle Data Group.\cite{PDG}
The current values from solar neutrinos are

\begin{equation}
\Delta m^2_{21} = (8.0\pm 0.3)\times 10^{-5}~\mathrm{eV}^2 \quad
\mathrm{and} \quad \sin^2(2\theta_{12}) = 0.86^{+0.03}_{-0.04}
\end{equation}

\noindent
at the 68\% confidence level.
The current bounds from atmospheric neutrinos are

\begin{equation} \label{eq1}
\Delta m^2_{32} = (1.9\ \mathrm{to}\ 3.0)\times 10^{-3}~\mathrm{eV}^2 \quad
\mathrm{and} \quad \sin^2(2\theta_{23}) > 0.92
\end{equation}

\noindent
at the 90\% confidence level.
The best fit is $\Delta m^2_{32} = 2.4\times 10^{-3}$~eV$^2$.
Only a bound is known on the third angle $\sin^2(2\theta_{13}) < 0.19$.

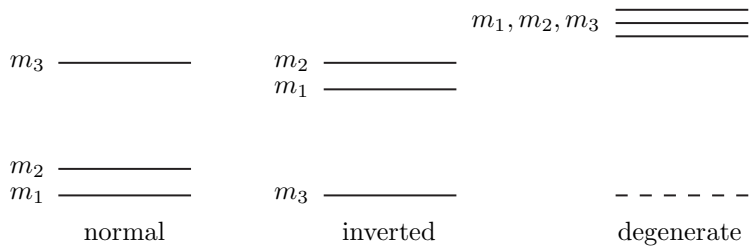
\begin{figure}[htb]
\begin{center}
\begin{picture}(290,105)(-25,-25)
\SetWidth{0.75}
%\Line(-25,-25)(265,-25)
%\Line(265,-25)(265,80)
%\Line(265,80)(-25,80)
%\Line(-25,80)(-25,-25)
\Line(0,50)(50,50)
\Line(0,10)(50,10)
\Line(0, 0)(50, 0)
\Text(-5,50)[r]{$m_3$}
\Text(-5,10)[r]{$m_2$}
\Text(-5,0)[r]{$m_1$}
\Text(25,-10)[tc]{normal}
\Line(100,50)(150,50)
\Line(100,40)(150,40)
\Line(100, 0)(150, 0)
\Text(95,50)[r]{$m_2$}
\Text(95,40)[r]{$m_1$}
\Text(95,0)[r]{$m_3$}
\Text(125,-10)[tc]{inverted}
\Line(210,70)(260,70)
\Line(210,65)(260,65)
\Line(210,60)(260,60)
\DashLine(210,0)(260,0){5}
\Text(205,65)[r]{$m_1, m_2, m_3$}
\Text(235,-10)[tc]{degenerate}
\end{picture}
\end{center}
\caption{\label{hierarchy}Possible neutrino mass hierarchies.}   
\end{figure}

The mass-squared differences can be accommodated within the SM of three active
neutrino flavours $\nu_e$, $\nu_\mu$, and $\nu_\tau$.
Since only the absolute difference of the squares of masses has been
determined in atmospheric neutrino oscillation experiments, the order of
the masses is not known.  
Possible mass hierarchies are shown in Fig.~\ref{hierarchy}.
For the normal and inverted hierarchies, we typically assume $\Delta
m^2_\mathrm{atm} = \Delta m^2_{32} \sim \Delta m^2_{31}$. 

For the normal and inverted mass hierarchies, the values of $m_1, m_2$,
and $m_3$ are assumed to be of the same order of magnitude as the larger
of the mass differences.
For the normal mass hierarchy, $m_1 = 0$ and $m_3 \gg m_2$ is usually
assumed. 
This leads to $m_3 = \sqrt{\Delta m^2_\mathrm{atm}} \approx 0.05$~eV
and $m_2 = \sqrt{\Delta m^2_{\astrosun}} \approx 0.009$~eV.
For the inverted mass hierarchy, $m_3 = 0$ and $m_1 \approx m_2$ is
usually assumed. 
This leads to $m_1 = m_2 = \sqrt{\Delta m^2_\mathrm{atm}} \approx
0.05$~eV.
In the degenerate mass scheme, $m_1 = m_2 = m_3 \approx 1$~eV is a
common arbitrary choice. 

Current cosmological data  and some cosmological assumptions suggest
that\cite{PDG}    

\begin{equation}
\sum_i m_i < (0.17 - 2.0)~\mathrm{eV}\, .
\end{equation}

\noindent
Here the sum is over neutrino states that were in thermal equilibrium
in the early universe. 
The mass of the heaviest neutrino is bounded by\cite{PDG}

\begin{equation}
0.04~\mathrm{eV} < m_\mathrm{heaviest} < (0.07 - 0.7)~\mathrm{eV}\, ,
\end{equation}

\noindent
where the lower bound is because the mass of the heaviest neutrino cannot
be lighter than $\sqrt{\Delta m_\mathrm{atm}^2}$.
This suggests that a more reasonable value to choose for the neutrino
mass in the degenerate scheme is 0.7~eV.

%%%%%%%%%%%%%%%%%%%%%%%%%%%%%%%%%%%%%%%%%%%%%%%%%%%%%%%%%%%%%%%%%%%%%%%%%%%%%%%
\section{Right-Handed Neutrinos in Large Extra Dimensions} \label{sec3}

In this section, we review the aspects of neutrinos in large extra
dimensions that will be needed to discuss their effects on Higgs
bosons.  
The ansatz of the large extra dimension paradigm is encompassed in the
following relationship.

\begin{equation} \label{eq5}
\bar{M}_\mathrm{Pl}^2 = \bar{M}_D^{2+\delta} V_\delta = M_D^{2+\delta}
R^\delta\, ,  
\end{equation}

\noindent
where $\bar{M}_\mathrm{Pl} = 1/\sqrt{8\pi G_N} \approx 2.4\times
10^{18}$~GeV is the reduced 4-dimensional Planck scale, $\delta$ is the 
number of additional spatial dimensions, $M_D$ is the fundamental Planck
scale of gravity in $(4 + \delta)$-spacetime dimensions, and $V_\delta =
L_1,\ldots,L_\delta$ is the volume of the compact extra-dimensional
space (where $L_i$ is the size of the $i$th compact dimension).
Assuming the volume has the configuration of a torus, $L_i = 2\pi R_i$.
In the simple case where all of the compact extra dimensions have equal
radii $R$, $V_\delta = (2\pi R)^\delta$. 
The case of non-equal radii and configurations other than a torus will
be discussed in section~\ref{sec3.4}. 

Throughout this paper, $M_D$ is used as the definition of the
fundamental Planck scale linking experimental measurements to the
theory. 
In the literature on bulk neutrinos in large extra dimensions, it is
common to use $M_*$ defined by $\bar{M}_\mathrm{Pl}^2 = M_*^{2+\delta}
R^\delta$, and consider $R^\delta$ as the volume in $\delta$-dimensional
space. 

Limits on $M_D$ and $R$ have been set by direct gravity measurements,
experiments at accelerators, and constraints from astrophysics and
cosmology. 
The E{\"o}t-Washington group constrained the size of the largest extra  
dimension to $R < 44~\mu$m at the 95\% confidence
level.\cite{Kapner07}
This completely rules out TeV-scale gravity with one large extra
dimension.
For two large extra dimensions, they obtain $R < 30~\mu$m.\cite{PDG}
The sensitivity to three or more extra dimensions of equal size is only
weakly constrained by accelerator experiments.
For $\delta = 3$, $M_D$ is greater than 1.2~TeV and for $\delta = 4$,
$M_D$ is greater than 0.94~TeV from the LEP experiments. 
For $5 \le \delta \le 8$, $M_D$ is greater than 0.8~TeV from the
Tevatron experiments. 
The astrophysical and cosmological limits on $M_D$ are high,
particularly for two or three extra dimensions.  
However, they are based on a number of assumptions so the results are
only order of magnitude estimates. 
Thus, we will not consider further astrophysical or cosmological limits.
We will often be interested in the quantity $1/R$.
The best limits on $1/R$ based on the numbers above are
$1/R > 7\times 10^{-3}$~eV for $\delta = 2$, 
$1/R > 75$~eV for $\delta = 3$, and 
$1/R > 2\times 10^4$~eV for $\delta = 4$.

When adding a right-handed neutrino to the paradigm of large extra
dimensions, there are two common approaches taken.
One approach is to introduce a separate bulk fermion for each
flavour of neutrino on the SM brane.\cite{Dvali99,Davoudiasl02}
This approach extends the concept of flavour into the bulk.
A second approach is to introduce only one bulk fermion and give
this fermion a flavour-universal coupling to all three SM brane
neutrinos.\cite{Dienes01,Lam01} 
In this approach, flavour is a feature internal to the SM and is
restricted to the SM brane.
For simplicity of illustration, we will consider a model of three
separate bulk neutrinos. 

We can view the Lagrangian density as split into separate bulk and SM
brane contributions.
The action can be written as

\begin{equation} \label{eq3}
S = \int d^4x d^\delta y \left[ \mathcal{L}_\mathrm{bulk} +
\delta(\vec{y}) \mathcal{L}_\mathrm{brane} \right]\, ,  
\end{equation}

\noindent
where $\mathcal{L}_\mathrm{brane}$ contains the usual SM Lagrangian plus
the interaction terms with the bulk fields, while
$\mathcal{L}_\mathrm{bulk}$ contains the dynamical terms for the bulk 
fields and the usual Einstein-Hilbert gravity term.
The extra $\delta$-dimensions are represented by the coordinates
$\vec{y} = (y^1, \ldots , y^\delta)$. 
Since we will only be considering thin SM branes, we use the delta
function $\delta(\vec{y})$ to precisely locate the SM fields and connect
the fields on the two manifolds. 
Small spreading from $y=0$, such as in models of split fermions, have
also been considered.\cite{Dienes06}  

To further simplify things, we will consider a 5-dimensional theory.
In addition, a theory of more than one extra dimension approximates to
the 5-dimensional theory when one of the extra dimensions is much larger 
than the sizes of the other dimensions. 
The generalisation to more than one extra dimension is straight forward
except for one non-trivial detail.
This is how the higher-dimensional spinor in the bulk couples to the
lepton spinor on the 3-brane.
For a generalisation of the 6-dimensional case see
Ref.~\refcite{Matias05}. 

We begin by adding Dirac fermions to the bulk with wave function
$\Psi^\alpha(x^\mu,\vec{y})$, where $\alpha = 1,2$, and 3 is the flavour 
index.
In the 4-dimensional Weyl basis, a 4-component Dirac spinor can
be decomposed as   

\begin{equation}
\Psi^\alpha(x^\mu,\vec{y}) = \left[
\begin{array}{c} 
\psi^\alpha_L(x^\mu,\vec{y})\\
\psi^\alpha_R(x^\mu,\vec{y})\\
\end{array}
\right]\, ,
\end{equation}

\noindent
where $\psi_L^\alpha$ and $\psi_R^\alpha$ are 2-component complex Weyl
spinors.   
The $L$ and $R$ subscripts explicitly indicate the 4-dimensional Lorentz 
property.
For the next little while, the flavour index is omitted to simplify
the notation. 

The 5-dimensional coordinates are $x^A \equiv (x^\mu,y)$, where $x^\mu$
are not compactified and $y$ is a single coordinate perpendicular to the 
brane. 
The $y$-direction is compactified on a circle of circumference $2\pi R$
by making the periodic identification $y \sim y + 2\pi R$.
On the 3-brane $y=0$.

In the Weyl basis, the 5-dimensional Dirac matrices $\Gamma$ have
size $4\times 4$ and can be written as 

\begin{equation}
\Gamma^\mu = \left(
\begin{array}{cc} 
0 & \sigma^\mu\\
\bar{\sigma}^\mu & 0\\
\end{array}
\right)
\quad \mathrm{and} \quad
\Gamma^5 = -i\gamma_5 = \left(
\begin{array}{cc} 
i & 0\\
0 & -i\\
\end{array}
\right)\, ,
\end{equation}

\noindent
where $\sigma^\mu = (1,\sigma^i)$ and $\bar{\sigma}^\mu =
(1,-\sigma^i)$, and the $\sigma^i$ are the three Pauli matrices. 

The free action for bulk fermions is

\begin{equation}
S_\mathrm{bulk} = \int d^4x dy \bar{\Psi}(x,y) i \Gamma^A \partial_A
\Psi(x,y)\, ,   
\end{equation}

\noindent
where

\begin{equation}
\bar{\Psi} = \Psi^\dag \Gamma^0
\quad \mathrm{and} \quad
\Psi^\dag = (\psi_L^\dag, \psi_R^\dag)\, .
\end{equation}

\noindent
The bulk action in terms of 2-component spinors is 

\begin{equation} \label{eq11}
S_\mathrm{bulk} = \int d^4x dy \left[ 
\psi_R^\dag i \sigma^\mu \partial_\mu \psi_R + \psi_L^\dag
i\bar{\sigma}^\mu \partial_\mu \psi_L
- \psi_R^\dag \partial_5 \psi_L + \psi_L^\dag \partial_5 \psi_R
\right]\, .
\end{equation}

To obtain the 4-dimensional effective theory, it is common to
compactify the large (flat) extra dimensions on a torus by making the
identification $y^i \sim y^i + 2\pi R$ for each dimension.
This suggests performing a KK expansion of the form

\begin{equation} \label{eq12}
\psi(x^\mu,\vec{y}) = \sum_{\vec{n}} \psi^{(\vec{n})}(x^\mu)
f_{\vec{n}}(\vec{y})\, , 
\end{equation}

\noindent
where $\vec{n} = (n_1,\ldots,n_\delta)$ is a vector in number space
($n_i$ can be positive, negative, or zero), $\psi^{(\vec{n})}$ are the
KK modes, and $f_{\vec{n}}(\vec{y})$ are a complete set of periodic
orthogonal functions over the $\vec{y}$ space that satisfy 

\begin{equation}
\int_0^{2\pi R} d^\delta y f^\dag_{\vec{n}}(\vec{y}) f_{\vec{m}}(\vec{y}) =
\delta_{\vec{n}\vec{m}}\, . 
\end{equation}

\noindent
The usual choice for $f_{\vec{n}}(\vec{y})$ that exhibits the wave
nature of the wave function is 

\begin{equation} \label{eq14}
f_{\vec{n}}(\vec{y}) = \frac{e^{-2\pi i\vec{n} \cdot \vec{y}
/(V_\delta)^{1/\delta}}}{\sqrt{V_\delta}}\, .
\end{equation}

\noindent
For higher-dimensional fermion fields, the KK expansion must include a
higher-dimensional Weyl spinor.
In the simplified 5-dimensional case considered here, the spinor is the
same as the usual 4-dimensional Weyl spinor and is contained in
$\psi^{(\vec{n})}$.   
For higher-dimensions, $\psi^{(\vec{n})}$ must have a vector structure
and the representation of the chirality multiplets is not unique.

Often Eq.~(\ref{eq14}) is used for the KK expansion and boundary
conditions are imposed afterwards.
Since the spectra resulting from toroidal compactification is typically
not chiral, we will compactify on an orbifold
$\mathcal{T^\delta}/\mathcal{Z}_2$:  
the quotient space of a $\delta$-dimensional torus will inversion
symmetry. 
When an extra space dimension is compactified under the group of
$\mathcal{Z}_2$ isometries, it is natural for one of the 2-component
Weyl spinors, e.g. $\psi_R$, to be taken to be even under the
$\mathcal{Z}_2$ action $y\to -y$, while the other spinor $\psi_L$ is
taken to be odd.
The left-handed SM neutrino $\nu_L$ is restricted to a brane localised 
at the orbifold fixed point $y = 0$, while $\psi_L$ vanishes at this point.
From the 4-dimensional point of view, a higher-dimensional SM singlet
fermion can be decomposed into a tower of KK excitations 

\begin{eqnarray}
\psi_R(x,y) & = & \frac{1}{\sqrt{2\pi R}} \psi_R^{(0)}(x) +
\frac{1}{\sqrt{\pi R}} \sum_{n=1}^\infty 
\psi_R^{(n)}(x) \cos\left(\frac{ny}{R}\right)\, ,\label{eq15}\\
\psi_L(x,y) & = & \frac{1}{\sqrt{\pi R}} \sum_{n=1}^\infty
\psi_L^{(n)}(x) \sin\left(\frac{ny}{R}\right)\, ,\label{eq16}
\end{eqnarray}

\noindent
where $\psi_{R/L}^{(n)}(x)$ are 4-dimensional states: right- and
left-handed Weyl spinors.  
The first term of $\psi_R$ is the zero mode and is independent of the
extra dimensions.
The other terms are called KK modes.
We point out that $\psi_L$  does not have a zero mode.  

The 4-dimensional effective theory is obtained by substituting
Eq.~(\ref{eq15}) and Eq.~(\ref{eq16}) into Eq.~(\ref{eq11}), and
integrating out $y$ (dimensional reduction) to obtain 

\begin{eqnarray}
S_\mathrm{bulk} & = & \int d^4x \left[ \psi_R^{(0)\dag}
  i\sigma^\mu \partial_\mu 
\psi_R^{(0)} + \sum_{\hat{n}=1}^\infty \left( \psi_R^{(\hat{n})\dag}
i\sigma^\mu \partial_\mu \psi_R^{(\hat{n})} + \psi_L^{(\hat{n})\dag}
i\bar{\sigma}^\mu \partial_\mu \psi_L^{(\hat{n})} \right) \right. \nonumber\\
& & \left. - \sum_{\hat{n}=1}^\infty
\frac{|\hat{n}|}{R} \left( \psi_R^{(\hat{n})\dag}
  \psi_L^{(\hat{n})} + \psi_L^{(\hat{n})\dag} \psi_R^{(\hat{n})} \right)
  \right]\, .   
\end{eqnarray}

\noindent
We have now generalized beyond a single extra dimension to the higher
extra-dimensional case.  
The vector in number space now has magnitude $|\hat{n}| = \sqrt{n_1^2 +
\cdots + n_\delta^2}$, where $n_i$ includes only the positive modes, but
excludes the zero mode. 
We observe the usual tower of KK states in which the Dirac masses are
$|\hat{n}|/R$.  
The sum over $\hat{n}$ can in principle go to infinity, leading to an infinite
tower of Dirac KK states.
However, we take the view that this extra-dimensional effective field
theory description is only valid up to a cut-off scale of approximately
$M_D$ and therefore truncate the sum such that the highest KK mass is
below $M_D$. 
The zero mode $\psi_R^{(0)}$ decouples from the tower of KK states and
is exactly massless.  

In the effective 4-dimensional theory, the most general $SU(2)$
invariant expression describing the interaction between brane and bulk
fields is    

\begin{eqnarray} \label{eq18}
S_\mathrm{int} & = & -\frac{g}{\sqrt{\bar{M}_D^\delta}} \int d^4x
\bar{\ell}_L(x) \phi_c(x) \psi_R(x,\vec{y}=0) + H.c.\nonumber\\   
& = & -\frac{g}{\sqrt{\bar{M}_D^\delta V_\delta}} \int d^4x 
\left( \nu^\dag_L, \ell^\dag_L \right) 
\left( \begin{array}{c} \bar{\phi}^0\\ -\phi^- \end{array} \right)
\left[ \psi_R^{(0)} + \sqrt{2} \sum_{\hat{n}=1}^\infty
\psi^{(\hat{n})}_R \right] + H.c.\, ,   
\end{eqnarray}

\noindent
where $\ell_L$ is a left-handed lepton doublet, $\phi_c$ is the SM Higgs
doublet with hypercharge $-1$, and $g$ is a dimensionless Yukawa
coupling constant.  
The flavour indices in Eq.~(\ref{eq18}) have been suppressed.
The coupling $g$ can be made diagonal in flavour space by applying two
unitary transformations (see section~\ref{sec3.1}).   
The lepton doublet and Higgs field lie on the 3-brane, while the
massless Dirac fermion $\psi_R$ propagates in the full extra-dimensional
space. 
The coupling of Eq.~(\ref{eq18}) breaks the full Poincar{\'e} invariance
of the theory by picking out the component $\psi_R$ from the full Dirac
spinor $\Psi$. 
This can be expected since the presence of the wall itself breaks the
higher-dimensional Poincar{\'e} transformations. 
The four dimensional theory is still Lorentz invariant.

One of the crucial questions in explaining neutrino masses and
oscillations is the violation of lepton number. 
We assume lepton number is conserved and assign $\Psi$ the opposite
lepton number to the lepton doublet in Eq.~(\ref{eq18}).
Since we do not include Majorana masses, the action conserves lepton
number and only Dirac neutrino masses are possible for the left-handed
neutrinos. 
There have been a number of models that combine the ideas of large extra
dimensions with additional ingredients, such as, small Majorana masses
for the brane neutrinos.\cite{Dienes99}  

The KK states do not mix with each other and the left-handed KK states
do not interact with the SM fields.
However, the KK states do not completely decouple from the system.
The lowest-lying active neutrino (SM neutrino $\nu_L$) 
will mix with the entire tower of $\psi_R^{(\hat{n})}$ states.
Thus the KK states can participate in neutrino oscillations, acting  
effectively as a large number of sterile neutrinos.

After $SU(2)$ symmetry breaking by the Higgs mechanism, 

\begin{equation}
\phi_c = \left( 
\begin{array}{c}
\frac{v+H(x)}{\sqrt{2}}\\ 0
\end{array}
\right)\, .
\end{equation}

\noindent
The Higgs field acquires a vacuum expectation value (VEV) $ v =
(\sqrt{2} G_F)^{-1/2} = m_W/(2g_W) = 246$~GeV§. 
The interaction term in the action becomes

\begin{equation} \label{eq17}
S_\mathrm{int} = -\frac{m_\mathrm{D}}{v} \int d^4x \bar{\nu}_L H \left[
\psi_R^{(0)} + \sqrt{2}  \sum_{\hat{n}=1}^\infty \psi_R^{(\hat{n})}
\right] + H.c.\, , 
\end{equation}

\noindent
where

\begin{equation} \label{eq21}
m_\mathrm{D} = \frac{g v}{\sqrt{2\bar{M}_D^\delta V_\delta}}
= \frac{g}{\sqrt{2}}
\frac{\bar{M}_D}{\bar{M}_\mathrm{Pl}} v\, . 
\end{equation}

\noindent
Thus, we see that the interactions between the bulk fermions and the
brane fields generate Dirac mass terms between the brane fields and all
the KK modes of the singlet neutrinos via their Yukawa coupling to the
Higgs VEV.  
The Yukawa couplings of the bulk fields are suppressed by the volume
of extra dimensions.
The last expression in Eq.~(\ref{eq21}) appears to be independent of
the number of extra dimensions; the number of extra dimensions is
hidden in the definition of $\bar{M}_D =
M_D/(2\pi)^{\delta/(2+\delta)}$.   
In terms of known values

\begin{equation}
m_\mathrm{D} \sim g \frac{\bar{M}_D}{\mathrm{1 TeV}} \times
10^{-4}~\mathrm{eV}\, ,  
\end{equation}

\noindent
and thus small neutrino masses consistent with neutrino flavour
oscillation experiments can be obtained.

In summary, active neutrinos in the SM are Weyl particles of
left-handed helicity. 
Due to gauge invariance, they have no bare mass term.
However, left-handed neutrinos on the brane couple to the bulk
right-handed fermions, and their interaction allows them to aquire a 
mass. 
The left-handed bulk fermions do not couple to particles on the brane
since their wave function vanishes at $y=0$, and thus they decouple from
the theory.

%%%%%%%%%%%%%%%%%%%%%%%%%%%%%%%%%%%%%%%%%%%%%%%%%%%%%%%%%%%%%%%%%%%%%%%%%%%%%%%
\subsection{Neutrino Mass} \label{sec3.1}

Next we examine the mass eigenvalues and eigenstates.
For each neutrino flavour, the mass matrix in KK space can be
diagonalised independently. 
Many authors have diagonalised the mass matrix and written down the
neutrino mass eigenstates in a variety of different basis. 
We follow the approach of Cao, Gopalakrishna and Yuan.\cite{Cao04a}

By collecting the neutrino mass terms in the Lagrangian and explicity
including the neutrino flavour indices $\alpha$ and $\beta$, we obtain

\begin{equation} \label{eq23}
\mathcal{L}_\mathrm{mass} = - \sum_{\alpha=1}^{3}
\sum_{\hat{n}=1}^{\infty} \frac{|\hat{n}|}{R} \psi_R^{\alpha(\hat{n})\dag}
\psi_L^{\alpha(\hat{n})} - \sum_{\alpha,\beta=1}^{3}
\frac{m_\mathrm{D}^{\alpha\beta}}{v} H \left( \psi_R^{\alpha(0)\dag} + \sqrt{2}
\sum_{\hat{n}=1}^{\infty} \psi_R^{\alpha(\hat{n})\dag} \right)
\nu_L^\beta + H.c.  
\end{equation}

\noindent
We make the Yukawa coupling diagonal in the flavour space by applying
the rotations\cite{Davoudiasl02}

\begin{equation}
\begin{array}{ccccccc}
\nu_L^\alpha & = & l^{\alpha i} \nu_L^{\prime\,i}, & \quad &
\nu_R^\alpha & = & \left( r^{\alpha i} \right)^* \psi_R^{\prime\,i (0)},\\
\psi_L^{\alpha (\hat{n})} & = & r^{\alpha i} \psi_L^{\prime\,i(\hat{n})},
& \quad & \psi_R^{\alpha (\hat{n})} & = & \left( r^{\alpha i} \right)^*
\psi_R^{\prime\,i (\hat{n})},\\ 
\ell_L^\alpha & = & l_\ell^{\alpha i} \ell_L^{\prime\,i}, & \quad &
\ell_R^\alpha & = & \left( r_\ell^{\alpha i} \right)^* \ell_R^{\prime\,i},
\end{array}
\end{equation}

\noindent
where the $3\times 3$ unitary matrices $l$ and $r$ are chosen to
diagonalise $m_\mathrm{D}^{\alpha\beta}$, such that $(r^{\alpha i})^*
m_\mathrm{D}^{\alpha\beta} l^{\beta j} = m_\mathrm{D}^i \delta^{ij}$.
The unitary matrices $l_\ell$ and $r_\ell$ are similarly chosen to
diagonalise the charged lepton mass matrix.  
This choice of matrices has an advantage in that it does not affect the
diagonality of the first term in Eq.~(\ref{eq23}).
The charged current interactions now become proportional to the PMNS
matrix\cite{Pontecorvo58}$^-$\cite{MNS}

\begin{equation}
V_\mathrm{PMNS} \equiv l_\ell^\dag l\, .
\end{equation}

The Dirac spinors can be define via

\begin{equation}
\nu \equiv \left( \begin{array}{c}\nu_L \\ \psi_R^{(0)}\end{array} \right),\
\nu^{(1)} \equiv \left( \begin{array}{c}\psi_L^{(1)} \\
\psi_R^{(1)}\end{array} \right),\ \cdots, 
\nu^{(\hat{n})} \equiv \left( \begin{array}{c}\psi_L^{(\hat{n})} \\
\psi_R^{(\hat{n})}\end{array} \right),\ \cdots\, . 
\end{equation}

\noindent
Also, for each neutrino flavour, the neutrino mass term in the Lagrangian
density can be written as

\begin{equation}
\mathcal{L}_\mathrm{mass} = \bar{\nu}_\mathrm{D} M \nu_\mathrm{D}\, ,
\end{equation} 

\noindent
where $\nu_\mathrm{D}^T = (\nu,\nu^{(1)},\cdots, \nu^{(\hat{n})},\cdots)$.
In this basis, the mass matrix is

\begin{equation}
M = \left( \begin{array}{cccccc}
m_\mathrm{D} & \sqrt{2}m_\mathrm{D} P_R & \cdots & \sqrt{2}m_\mathrm{D}
P_R & \cdots\\ 
\sqrt{2}m_\mathrm{D} P_L & 1/R & \cdots & 0 & \cdots\\
\vdots & \vdots & \ddots & \vdots & \cdots\\
\sqrt{2}m_\mathrm{D} P_L & 0 & \cdots & |\hat{n}|/R & \cdots\\
\vdots & \vdots & \vdots & \vdots & \ddots\\
\end{array} \right)\, ,
\end{equation}

\noindent
where $P_{L/R} \equiv (1 \mp \gamma_5)/2$ are the usual chiral projection
operators.
This mass matrix is infinite dimensional.
The mode number $|\hat{n}|$ is degenerate with degeneracy $d_n$, and the 
$|\hat{n}|/R$ elements in the mass matrix are $d_n\times d_n$ block
diagonal.   

The mass matrix can be diagonalized by two unitary matrices such that
$\nu_D = (L P_L + R P_R) \tilde{\nu}_D$, where $\tilde{\nu}_D$ is the
mass eigenvector.  
We can reintroduce the generation index $i$ and the KK index $n$ to
write the flavour state $\nu_L^\alpha$ in terms of the mass eigenstates
$\tilde{\nu}_L^{i(n)}$ as  

\begin{equation} \label{eq24}
\nu_L^\alpha = l^{\alpha i} L_i^{0 n} \tilde{\nu}_L^{i(n)}\, ,
\end{equation}

\noindent
where $L_i^{0 n}$ is the first row of the $L_i$ unitary matrix.

If we treat the off-diagonal terms in the mass matrix as
perturbations\footnote{The strong coupling limit has also been
investigated.\cite{Lam01}}, then the zeroth-order lightest eigenvalue
of the mass matrix is $m_\mathrm{D}$. 
In the limit 

\begin{equation} \label{eq25}
(m_\mathrm{D} R)^2 \sum_{\hat{n}} \frac{d_n}{|\hat{n}|^2} \ll 1\, , 
\end{equation}

\noindent
we have, to a good approximation, a Dirac fermion $\nu$ with mass
$m_\mathrm{D}$, and additional Dirac fermions $\nu^{(\hat{n})}$ with
masses $|\hat{n}|/R$.  
If the neutrino's mass is less than $1/R$, the massive KK modes will
have little effect on the neutrino mass term and only the zero mode KK
state will generate neutrino mass. 
In higher order, the lowest mass eigenstate gets an admixture of masses
from the KK modes $\psi_R^{(\hat{n})}$.
(Note that only the right-handed components of the KK states mix with
the SM neutrino). 
The perturbative limit in Eq.~(\ref{eq25}) is satisfied when the brane to
bulk coupling $m_\mathrm{D}$ is small compared to the compactification
scale $1/R$. 
It is also necessary to satisfy Eq.~(\ref{eq25}) so that the probability
of active neutrinos oscillating into the sterile KK states is small.  

Adding the second order correction to the lowest mass eigenvalue
$m_\mathrm{D}$ gives

\begin{equation} \label{eq26}
m_\nu = m_\mathrm{D} \left[ 1 - (m_\mathrm{D} R)^2 \sum_{\hat{n}}
\frac{d_n}{|\hat{n}|^2} \right]\, ,
\end{equation}

\noindent
where we identify $m_\nu$ as the physical SM neutrino mass.
For convenience, we define $m_\nu = m_\mathrm{D}/N$ and proceed to
calculate the correction $N$\footnote{Some authors have defined the
same normalisation as $\sqrt{N}$.}.
This is also the normalisation of the wave function of the lowest
eigenstate of the mass matrix. 
For a single extra dimension, $\hat{n} = n_1 = n$ and there is no
degeneracy in the mass matrix ($d_n=1$).  
For $\delta >1$, the states with mass $|\hat{n}|/R$ can be degenerate
with degeneracy $d_n$ at the $\hat{n}$th level.
For a large number of KK modes (also possible large $|\hat{n}|$), we can
replace the sum over modes by an integral.
The leading behaviour is given by the surface of a $(\delta-1)$-sphere
of radius $|\hat{n}|$ in number space.\cite{Cao04a}
Hence, 

\begin{equation}
d_n \approx S_{\delta-1} |\hat{n}|^{\delta-1}\, ,
\end{equation}

\noindent
where $S_{\delta-1} = 2\pi^{\delta/2}/\Gamma(\delta/2)$ and $\Gamma$ is
the usual Euler gamma function.
In calculations, we sum the number of KK modes up to some maximum.
In this case, the heaviest KK state that could be produced is limited by
$M_D$. 
We denote $N_\mathrm{cut}$ to be the radius of the biggest sphere in
number space such that $N_\mathrm{cut}/R = M_D$. 
The sum over KK states can be divergent and depends on $N_\mathrm{cut}$. 
One can show (see \ref{appendA}) that

\begin{equation} \label{eq29}
N^2 = 1 + \left( \frac{m_\mathrm{D}}{M_D} \right)^2 \left(
\frac{\bar{M}_\mathrm{Pl}}{M_D} \right)^2 \times \left\{ \begin{array}{lc} 
\frac{\pi^2}{6} \left( \frac{\bar{M}_\mathrm{Pl}}{M_D} \right)^2 &
\mathrm{for}\ \delta = 1,\\ 
2\pi \ln \left( \frac{\bar{M}_\mathrm{Pl}}{M_D}\right)& \mathrm{for}\ 
\delta = 2,\\
\frac{2\pi^{\delta/2}}{\Gamma(\delta/2)} \frac{1}{\delta-2} & 
\mathrm{for}\ \delta > 2. \end{array}\right.
\end{equation}

\noindent
The sum over KK modes converges for $\delta=1$, is logarithmic
divergent for $\delta=2$, and is power divergent for $\delta > 2$.

The correction $N$ causes the physical neutrino mass to be bounded from above.
The upper bound depends on the characteristics of the higher-dimensional
space: $M_D$ and $\delta$. 
The physical mass $m_\nu$ does not have an extrema (besides 0) but
asymptotically approaches the maximum value 

\begin{equation} \label{eq30}
m_\nu^\mathrm{max} \approx \frac{M_D^2}{\bar{M}_\mathrm{Pl}} \times
\left\{ \begin{array}{lc}
\frac{\sqrt{6}}{\pi} \frac{M_D}{\bar{M}_\mathrm{Pl}} &
\mathrm{for}\ \delta = 1,\\
\frac{1}{\sqrt{2\pi\ln\left(\frac{\bar{M}_\mathrm{Pl}}{M_D}\right)}} &
\mathrm{for}\ \delta = 2,\\
\sqrt{\frac{\Gamma(\delta/2) (\delta-2)}{2\pi^{\delta/2}}} &
\mathrm{for}\ \delta > 2,\\
\end{array} \right.
\end{equation}

\noindent
when $m_\mathrm{D}\to \infty$.
However, values of $m_\mathrm{D}$ that are too high violate
Eq.~(\ref{eq25}). 
The mixing between the lightest neutrino of mass $m_\mathrm{D}$ and
heavier neutrinos introduces a correction to the physical neutrino mass
$m_\nu$, 

\begin{equation} \label{eq31}
m_\nu \approx \frac{m_\mathrm{D}}{\sqrt{1 + \left(
\frac{m_\mathrm{D}}{m_\nu^\mathrm{max}} \right)^2}}\, . 
\end{equation}

\noindent
Since there is some uncertainty in where exactly the divergence
(for $\delta >1$) in the sum over KK modes should be cut off,
$m_\nu^\mathrm{max}$, and thus $m_\nu$, is a bit uncertain.
However, $m_\nu$ must satisfy $m_\nu < m_\nu^\mathrm{max} <
m_\mathrm{D}$. 
Only a rigorously formulated theory, such as string theory, could give
precise knowledge of the cut-off parameter.

\begin{figure}[htb]
\begin{center}
\includegraphics[width=\columnwidth]{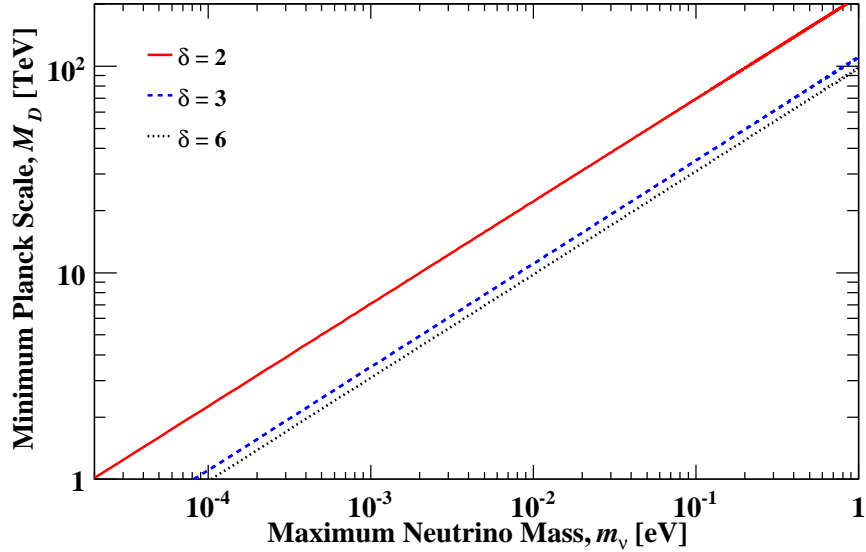}
\caption{\label{max}Minimum fundamental Planck scale $M_D$ versus maximum
neutrino mass $m_\nu$ required by the perturbative expansion of the
mass matrix for two, three, and six extra dimensions.}
\end{center}
\end{figure}

The perturbative condition places constraints on $m_\mathrm{D}$,
$\delta$, and $R$ (and $M_D$ due to the cut-off in the summation). 
Using Eq.~(\ref{eq5}) we can re-express the condition on $R$ in terms of 
$M_D$.
Since $m_\mathrm{D}$ is not physical, we can replace it by a function of
$m_\nu$ by using Eq.~(\ref{eq31}) or replace it by a function of $g$ by
use Eq.~(\ref{eq21}). 
We discuss the former case next, and then the later case in
section~\ref{sec3.3}.  
For one extra dimension and $M_D\sim 1$~TeV, the neutrino mass must be
less than about $10^{-19}$~eV.
In the same scenario, to obtain a neutrino mass of 1~eV, $M_D
\gtsimeq 1000$~TeV.
We find this small neutrino mass or large fundamental Planck scale to be 
unnatural and do not consider one extra dimension.
For $\delta > 1$, satisfying Eq.~(\ref{eq25}) gives the results shown
in Fig.~\ref{max}.   
Current lower limits on the fundamental Planck scale are consistent
with current upper limits on the neutrino mass for $\delta >1$.
For two extra dimensions, rather large values for the fundamental Planck 
scale and low values for the neutrino mass are required to satisfy the
perturbative condition. 
This makes $\delta = 2$ an unlikely choice in this model.
A value of the maximum neutrino mass given by the current limits on
atmospheric mixing data of about 0.05~eV would require the fundamental
Planck scale to be above about 20~TeV, for $\delta > 2$.
In section~\ref{sec3.4}, we will discuss a slight modification of the
model which relaxes these constraints.

%%%%%%%%%%%%%%%%%%%%%%%%%%%%%%%%%%%%%%%%%%%%%%%%%%%%%%%%%%%%%%%%%%%%%%%%%%%%%%%
\subsection{Neutrino Mixing} \label{sec3.2}

We now briefly consider the implications of higher dimensions on
neutrino oscillations.
The flavour eigenstates are a non-trivial combination of the physical
propagating energy eigenstates that cause the flavour eigenstates to
oscillate as a function of time.
Good fits to the data from oscillation experiments are obtained by
considering only three active species of neutrinos.
Strong constraints exist on the mixing of an active neutrino species to
a sterile neutrino species. 

Let $P_{\nu_\alpha \to \nu_\beta}$ be the probability of an active
neutrino species $\alpha$ oscillating into another active neutrino
species $\beta$ after travelling a distance $L$. 
The state of the neutrino after a time $t$ is determined by the
Hamiltonian $H$, and given by the time evolution operator $e^{-iHt}$.
Thus 

\begin{equation}
P_{\nu_\alpha \to \nu_\beta} = \left| \langle \nu_L^\beta \left|
e^{-iHt} \right| \nu_L^\alpha \rangle \right|^2\, . 
\end{equation}

Using Eq.~(\ref{eq24}) and allowing the Hamiltonian to act on the energy
eigenstate gives

\begin{equation}
P_{\nu_\alpha \to \nu_\beta} = \left| l^{\beta i *} l^{\alpha i} \left|
L_i^{0 n} \right|^2 d_ne^{-iE_i^{(n)}L} \right|^2\, ,
\end{equation}

\noindent
where $E_i^{(n)}$ is the energy eigenvalue of the $n$th KK mode of the
$i$th species. 

For a neutrino beam with energy $E_\nu$ and momentum $p_\nu$, 
$E_i^{(n)} \approx p_\nu + \left( m_i^{(n)} \right)^2 / (2E_\nu)$ in the
relativistic limit and $m_i^{(n)}$ are the mass eigenvalues.
Hence

\begin{equation} \label{eq39}
P_{\nu_\alpha \to \nu_\beta} = \left| l^{\beta i *} l^{\alpha i} \left( \left|
L_i^{0 0} \right|^2 e^{-i(L/2E_\nu)m_i^2} + \left| L_i^{0 \hat{n}}
  \right|^2 d_n e^{-i(L/2E_\nu)(m_i^{(\hat{n})})^2} \right) \right|^2\, . 
\end{equation}

The elements of the unitary matrix $L_i$ were given in
Ref.~\refcite{Cao04a}. 
Substituting these matrix elements into Eq.~(\ref{eq39}), summing over
$\beta$, and assuming $m_{Di}R/\hat{n} \ll 1$ gives

\begin{equation}
\sum_\beta P_{\nu_\alpha \to \nu_\beta} = 1 - 8 \left| l^{\alpha i} \right|^2
\xi_i^2 \sum_{\hat{n}} \frac{d_{\hat{n}}}{\hat{n}^2} \sin^2 \left(
\frac{L \hat{n}^2}{4E_\nu R^2} \right)
\end{equation}

\noindent
to second order in $m_{Di}R$.
We see that the oscillations consists of the interference of an infinite
number of modes with increasing frequency $\propto \hat{n}^2$ and
decreasing amplitudes $\propto 1/\hat{n}^2$.
In practice, the high frequency modes can be averaged and only a few low
frequency oscillations can be observed, depending on the energy
resolution of the detector.

The probability for an active neutrino state to oscillate into sterile
neutrino states $\nu_s$ is 

\begin{equation} \label{eq41}
P_{\nu_\alpha \to \nu_s} = 1 - \sum_\beta P_{\nu_\alpha \to \nu_\beta} =
8 \left| l^{\alpha i} \right|^2 \xi_i^2 \sum_{\hat{n}}
\frac{d_{\hat{n}}}{\hat{n}^2} \sin^2 \left( \frac{L\hat{n}^2}{4E_\nu
  R^2} \right)\, .  
\end{equation}

We can use the CHOOZ\cite{CHOOZ} and atmospheric neutrino data along
with Eq.~(\ref{eq41}) to set bounds on $1/R$.
The mass values given in section~\ref{sec2} can be used for each mass
hierarchy scheme. 
We work in a basis in which the charged lepton mass matrix is diagonal.  
In this case $V_\mathrm{PMNS} = l$.
For simplicity, we take $\theta_{13} = 0$ and the mixing for atmospheric
neutrinos to be maximal with $\theta_{23} = \pi/4$.
These simplifications give

\begin{equation} \label{eq42}
l = 
\left(
\begin{array}{ccc}
c_{12} & s_{12} & 0\\
-s_{12}/\sqrt{2} & c_{12}/\sqrt{2} & 1/\sqrt{2}\\
s_{12}/\sqrt{2} & -c_{12}/\sqrt{2} & 1/\sqrt{2}
\end{array}
\right) = 
\left(
\begin{array}{rrc}
0.829 & 0.559 & 0\\
-0.396 & 0.586 & 0.707\\
0.396 & -0.586 & 0.707
\end{array}
\right)\, .
\end{equation}

\noindent
The resulting limits\cite{Cao04a} are given in Table~\ref{oss}.
For $\delta >3$, the active state would oscillate mostly to the heaviest
states and we are not able to reliably estimate the oscillation
probability.
We take these constrains into consideration throughout this paper.

%%%%%%%%%%%%%%%%%%%%%%%%%%%%%%%%%%%%%%%%%%%%%%%%%%%%%%%%%%%%%%%%%%%%%%%%%%%%%%
\subsection{Neutrino Coupling} \label{sec3.3}

Now consider the Higgs interaction term in the action of Eq.~(\ref{eq17}).
The right-handed neutrino does not carry any electroweak quantum numbers
and therefore can be produced only through Yukawa interactions.
The Higgs boson couples to a tower of KK neutrino states with the same
Yukawa coupling ($\sim m_\mathrm{D}/v$).
In general, the production cross sections and decay widths will be
suppressed by the small Yukawa coupling (since the neutrino mass is
small), but enhanced by the sum over a large number ($\sim
10^{30/\delta}$) of KK excitations of bulk fermions. 

\begin{figure}[htb]
\begin{center}
\includegraphics[width=\columnwidth]{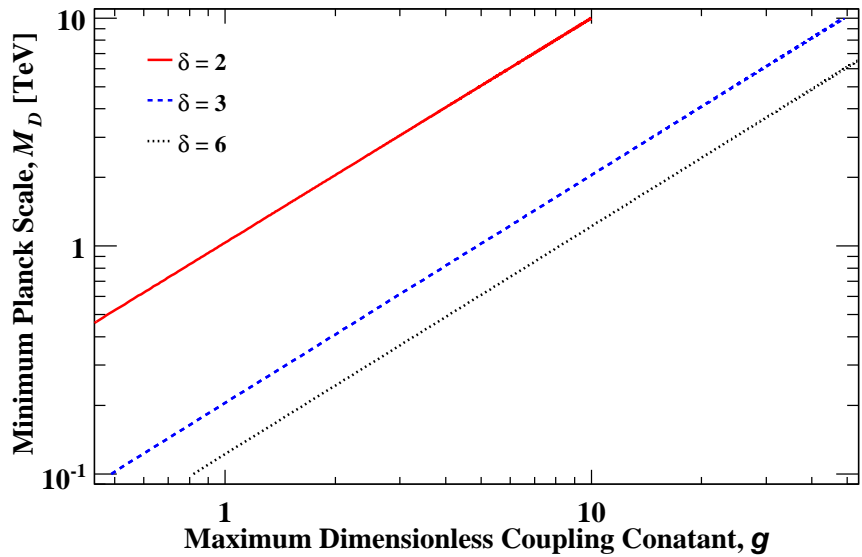}
\caption{\label{g}Minimum fundamental Planck scale $M_D$ versus maximum
dimensionless coupling constant $g$ required by the perturbative
expansion of the mass matrix for two, three, and six extra dimensions.} 
\end{center}
\end{figure}

As mentioned in the previous section, the perturbative condition
Eq.~(\ref{eq25}) can also be expressed in terms of $M_D$, $\delta$, and
$g$. 
Equation~(\ref{eq21}) can be substituted into Eq.~(\ref{eq25}), and a
condition on $M_D$ and $g$ can be obtained for a given $\delta$. 
For $\delta = 1$, $M_D$ is required to be close to the GUT scale of
$\mathcal{O}(10^{16})$~GeV or $g$ is required to be infinitesimally small. 
Thus a single extra dimension is not of interest to us.
For $\delta = 2$, $M_D/g \gtsimeq 1$~TeV, and for $\delta>2$, $M_D/g
\gtsimeq v$.
Both of these conditions should always be satisfied. 
Thus $M_D \gtsimeq 1$~TeV and $g \ltsimeq 1$ will satisfy the
perturbative constraint for all $\delta > 1$.
Some cases for specific values $\delta$ are shown in Fig.~\ref{g}.

\begin{figure}[htb]
\begin{center}
\includegraphics[width=\columnwidth]{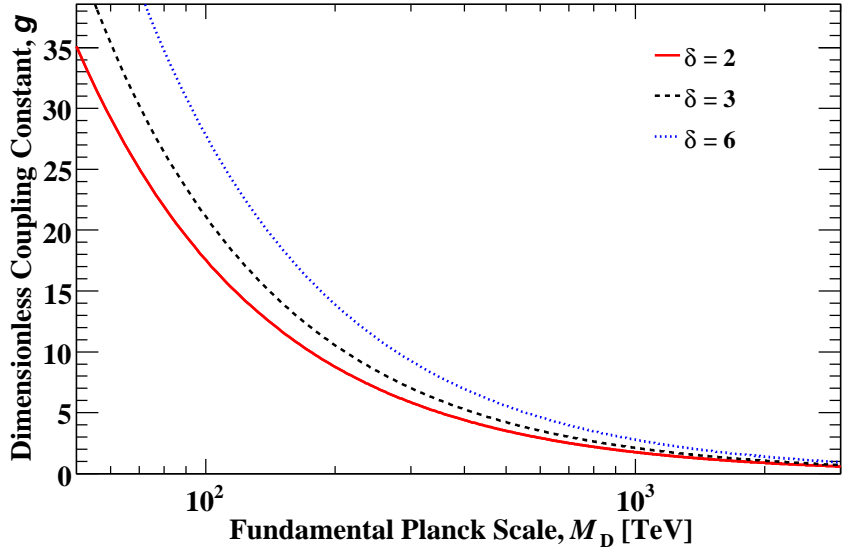}
\caption{\label{lambda}Dimensionless coupling constant $g$ versus
fundamental Planck scale $M_D$ for neutrino mass $m_\nu = 0.05$~eV and
2, 3, and 6 extra dimensions.} 
\end{center}
\end{figure}

Equations~(\ref{eq21}), (\ref{eq30}), and (\ref{eq31}) allow us to
relate $g$ to the physical neutrino mass $m_\nu$, $M_D$, and $\delta$. 
Based on the existing atmospheric neutrino mass best fit of $m_\nu =
0.05$~eV, Fig.~\ref{lambda} shows a plot of the dimensionless coupling
constant versus fundamental Planck scale. 
For a perturbative effective theory, $g$ can not be arbitrarily
large.\cite{Agashe01} 
For $g\sim 1$, $M_D$ must be about 2000~TeV.
This condition has very little dependence on the number of dimensions.
If we restrict $g < 10$, the fundamental Planck scale is required to be
$M_D > 200-300$~TeV.  
Since the UV cut-off scale for the KK sum is somewhat uncertain, $g \sim
\mathcal{O}(10)$ is not an unreasonable possibility, but will violate
the perturbative constraint for low $M_D$.
Thus, for the model of large extra dimensions that we have just
considered, $M_D$ is required to be quite large.
Hence we consider small modifications to the model in the next
subsection. 

%%%%%%%%%%%%%%%%%%%%%%%%%%%%%%%%%%%%%%%%%%%%%%%%%%%%%%%%%%%%%%%%%%%%%%%%%%%%%%%
\subsection{Bulk Fermions in Subspaces and Compactification} \label{sec3.4}

A bulk fermion may not necessarily propagate in the same $\delta$
extra-dimensional space that the graviton propagates in.
It is possible that the bulk fermion propagates in a subset
$\delta_\nu$ of the $\delta$-extra dimensions ($\delta > \delta_\nu$).
In fact, different bulk fermions could propagate in different dimensions
or the UV cut-off scale could be different for each.
Thus the formalism for generating small Dirac neutrino masses that we
have discussed so far is merely the specific case of $\delta =
\delta_\nu$. 
Assuming all the extra dimensions are the same size, the necessary
substitutions in the above formula are 

\begin{equation}
\delta \to \delta_\nu \quad \mathrm{followed\ by} \quad
\frac{\bar{M}_{Pl}}{M_D} \to \left( \frac{\bar{M}_{Pl}}{M_D}
\right)^{\delta_\nu/\delta}\, .  
\end{equation}

\noindent
The generalisation of Eq.~(\ref{eq29}) is given in \ref{appendA}.
In this model, the maximum value of the physical neutrino mass
Eq.~(\ref{eq30}) is given by 

\begin{equation}
m_\nu^\mathrm{max} \approx M_D \left( \frac{M_D}{\bar{M}_{Pl}}
    \right)^{\delta_\nu/\delta} \sqrt{\frac{\Gamma(\delta_\nu/2) 
    (\delta_\nu-2)}{2\pi^{\delta_\nu/2}}}
\end{equation}

\noindent
for the case of $\delta > 2$.
This is larger than the $\delta_\nu = \delta$ case.
The $\delta = 1$ and 2 cases follow trivially.

\begin{figure}[htb]
\begin{center}
\includegraphics[width=\columnwidth]{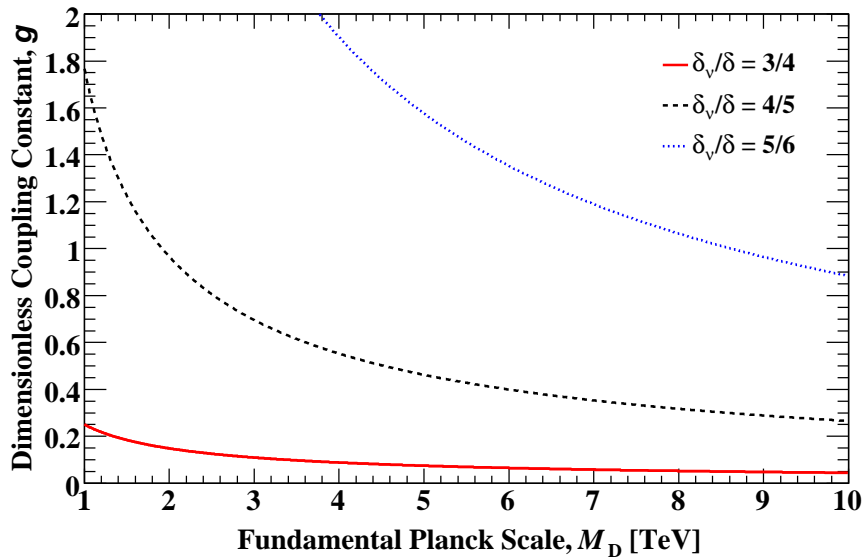}
\caption{\label{lambda-sub}Dimensionless coupling constant $g$ versus
fundamental Planck scale $M_D$ for three different neutrino subspaces of
gravity space.}  
\end{center}
\end{figure}

Although $\delta = 1$ is ruled out in the cases of gravity and in the
model just presented, $\delta_\nu = 1$ is not experimentally
constrained.  
However, for bulk fermions in one dimension, $m_\mathrm{D}R \ll 1$ will not be
satisfied.
Figure~\ref{lambda-sub} shows some cases for a bulk fermion living in a
subspace of gravity.
$M_D \sim\mathcal{O}(1)$~TeV is now possible for $g\sim\mathcal{O}(1)$. 

There is no reason for the internal $\delta$-dimensional manifold to be
symmetric.
One could imagine compactifying on a product of different dimensional
tori $\mathcal{T}^\delta = \mathcal{S}^1 \times \mathcal{S}^1 \times
\cdots \times \mathcal{S}^1$, each with its own characteristic
radius $R_i$. 
The compactification volume would then be $V_\delta = (2\pi)^\delta
R_1 R_2 \ldots R_\delta$, and the mass of the KK states would be

\begin{equation}
\sqrt{\sum_i \frac{n_i^2}{R_i^2}}\, .
\end{equation}

\noindent
We can create a simplified version of this scenario by adding three
assumptions to the theory.
These assumptions are that the bulk fermions propagate in a
sub-dimensional space of extra dimensions $\delta_\nu$ with common size
$R$, that gravity propagates in the space of extra dimensions $\delta$,
and that the extra dimensions 
($\delta-\delta_\nu$) have a common size $r$ with $r \ll R$.
With these simplifications Eq.~(\ref{eq5}) becomes

\begin{equation}
\bar{M}_\mathrm{Pl}^2 \sim \bar{M}_D^{\delta+2} (2\pi)^\delta R^{\delta_\nu}
r^{(\delta-\delta_\nu)}\, .
\end{equation}

\noindent
In this model, the Dirac mass for the SM neutrino Eq.~(\ref{eq21})
becomes  

\begin{equation}
m_\mathrm{D} \approx \frac{g}{\sqrt{2}}
\frac{1}{\sqrt{(2\pi\bar{M}_DR)^{\delta_\nu}}} v\, . 
\end{equation}

The volume in Eq.~(\ref{eq21}) is determined by the compactification
scheme.
The $(\pi R)^{\delta/2}$ factor in the denominator is identical if
compactified on a torus (circle) or a $\mathcal{Z}_2$ orbifold.
However, more interesting scenarios result in different volumes. 
For example, for a $\mathcal{Z}_N$ orbifold the volume becomes $(2\pi
R/N)^{\delta/2}$. 

By including such minor modifications as subspace or alternative
compactification schemes to the model, most constraints on the model can
be avoided.  
Throughout this paper we will use ``rectangular'' toroidal geometry for
compactification and express the results in terms of the fundamental
Planck scale by using Eq.~(\ref{eq5}).
A pedantic review of the subject would express all results in terms of
the compactification volume and leave it to the reader to choose the 
compactification scheme.  

We briefly point out yet another consideration in compactification.
Compactification manifolds are not only described by their volume but
also by their shape.
For example, a general 2-torus is described by three parameters:
$R_1$, $R_2$, and $\theta$.
Here $\theta$ is a shift angle that is usually taken to be $\pi/2$ for a
``rectangular'' torus. 
The shape parameters of the general 2-torus are $R_2/R_1$ and $\theta$.
The physical significance of the angle $\theta$ is that translations
along the $R_2$ direction produce simultaneous translations along the
$R_1$ direction.\cite{Dienes02a}
Both the volume and shape are important to fully describe the geometry
of the compactified extra dimensions.
It has been shown that the shape of compactification can dramatically
modify the KK spectrum regardless of whether the volume is changed or
not.\cite{Dienes02a}$^-$\cite{Dienes02c} 
Thus, the compactification scheme would not only affect the
phenomenology of models but also their interpretation if extra
dimensions were to be discovered.

%%%%%%%%%%%%%%%%%%%%%%%%%%%%%%%%%%%%%%%%%%%%%%%%%%%%%%%%%%%%%%%%%%%%%%%%%%%%%%%
\subsection{Constraints on the Size and Number of Extra Dimensions}

To end this section, we briefly discuss some constrains on models of
bulk neutrinos in large extra dimensions.
Since gravitons at low energies effect SM processes only marginally, the
existence of bulk neutrinos generally impose tighter constrains than
those due to gravitational interactions on the scale $M_D$. 
In the following, we discuss only a few of the many constraints on bulk
neutrinos that have been studied.  
We refer the reader to the literature for constraints due to 
anomalous magnetic moments of leptons;\cite{McLaughlin00} 
nuclear $\beta$-decay in nuclei;\cite{McLaughlin01}
rare charged lepton processes;\cite{Gouvea02} 
flavour-violating and universality-breaking phenomena involving $W$ and
$Z$ bosons, in addition to leptons;\cite{Ioannisian00} 
perturbative unitarity violation in Higgs-Higgs scattering;\cite{Cao04a}
$B$ meson and top quark decays through a virtual charged Higgs boson and
lepton flavour violating decays through a virtual charged Higgs
boson;\cite{Agashe00} 
neutral pion and neutral $B$ meson decays to invisible decay
modes;\cite{Deshpande03} 
and cosmological data.\cite{Abazajian03}

The luminosity from Supernova 1987a gives strong constraints on extra
dimensions.\cite{Barbieri00,Cacciapaglia03} 
The possible energy loss rate from SN1987a into invisible channels such
as the energy carried away by a large number of KK states leads to the
following restriction on the size of extra dimensions.
The maximum radius of any dimension is $1/R > 10$~keV or $R \ltsimeq
1$~{\AA}.  
The limit is independent of the number of dimensions.
For $\delta_\nu = \delta = 3$, this constraint requires $M_D \gtsimeq
20$~TeV. 

A constraint can also be obtained from big bang nucleosynthesis by the
need to avoid too much energy being dissipated into bulk KK neutrino
modes before the time of nucleosynthesis.
This leads an unacceptable expansion rate of the universe. 
For $\delta_\nu=\delta=2$, it is likely that too many of the heavy KK modes
would be thermal during
nucleosynthesis.\cite{Dvali99,Arkani01,Barbieri00,Goh02}
This would pose a problem for the expansion of the universe.

Intergenerational mass splitting and mixing in large extra dimensions
could lead to the violation of lepton universality and flavour changing
processes in the charged lepton sector.\cite{Faraggi99}
The mixing of a left-handed neutrino with heavy KK modes alters the tau
branching ratio and the decay widths of the charged pion, muon, and tau.
Based on experimental limits on lepton universality and flavour changing
transitions, $\bar{M}_D$ is required to greater than about 10~TeV for
most of the interesting range of neutrino mass
splitting.\cite{Faraggi99} 

There are also constraints on the radius of the extra dimensions
resulting from data obtained by oscillation experiments.
The data restrict the probability of an active neutrino state to mix
into a large number of sterile KK bulk neutrino states.   
Ref.~\refcite{Davoudiasl02} and Ref.~\refcite{Cao04a} derive the bounds
shown in Table~\ref{oss}.
Atmospheric neutrino measurements provide the most stringent bounds in
the normal mass hierarchy, while CHOOZ\cite{CHOOZ} data provides the
most stringent bounds for the inverted and degenerate mass hierarchies.   
Ref.~\refcite{Davoudiasl02} uses a model with three active brane
neutrinos and three bulk neutrinos.
In this case, the results are to be interpreted as constraints on the
size of the largest of the extra dimensions, regardless of their total
number. 

\begin{table}[htb]
\tbl{Lower bounds on $1/R$ (eV) inferred from oscillation experiments
for different mass hierarchies.
The results in the first row are from
Ref.~\protect\refcite{Davoudiasl02} and constrain the largest extra
dimension.
The results for 1, 2, 3 extra dimensions are from
Ref.~\protect\refcite{Cao04a}.}    
{\begin{tabular}{@{}cccc@{}} \toprule
Extra Dimensions & Normal & Inverted & Degenerate\\ \colrule
NA & 0.24 & 0.60 & 10.9\\ \colrule
1 & 0.15 & 0.5 & 10.6\\
2 & 1.5 & 5.3 & 100\\
3 & $5.6\times 10^3$ & $1.2\times 10^4$ & $10^5$\\
\botrule
\end{tabular} \label{oss}}
\end{table}

Given the above constraints and to avoid problems with standard big bang
nucleosynthesis, we require $\delta > 2$ throughout this paper.  

%%%%%%%%%%%%%%%%%%%%%%%%%%%%%%%%%%%%%%%%%%%%%%%%%%%%%%%%%%%%%%%%%%%%%%%%%%%%%%%
\section{Collider Signatures}

Testing the origin of small neutrino masses at the LHC is a hot and
important topic.\cite{Xing08,Xing09}
There are already many papers on how to test a variety of seesaw
mechanisms\cite{Aguila09} and look for heavy Majorana\cite{Atre09} (or
Dirac) neutrinos at the LHC.

\begin{figure}[htb]
\begin{center}
\begin{picture}(150,140)(-75,-70)
\SetWidth{0.75}
%\Line(-75,-70)( 75,-70)
%\Line( 75,-70)( 75, 70)
%\Line( 75, 70)(-75, 70)
%\Line(-75, 70)(-75,-70)
\ArrowLine(-50,0)(0,0)
\Vertex(0,0){1.5}
\DashLine(0,0)(50,50){5}
\ArrowLine(0,0)(50,-50)
\Text(-55,0)[r]{$\nu_L$}
\Text(55,50)[lb]{$H$}
\Text(55,-50)[lt]{$\nu_R$}
\end{picture}
\begin{picture}(150,140)(-75,-70)
\SetWidth{0.75}
%\Line(-75,-70)( 75,-70)
%\Line( 75,-70)( 75, 70)
%\Line( 75, 70)(-75, 70)
%\Line(-75, 70)(-75,-70)
\ArrowLine(-50,0)(0,0)
\Vertex(0,0){1.5}
\DashLine(0,0)(50,50){5}
\ArrowLine(0,0)(50,-50)
\Text(-55,0)[rb]{$\ell^{\;-}$}
\Text(55,50)[lb]{$H^-$}
\Text(55,-50)[lt]{$\nu_R$}
\end{picture}
\end{center}
\caption{\label{couple2}Higgs boson production involving right-handed bulk
neutrinos.}   
\end{figure}
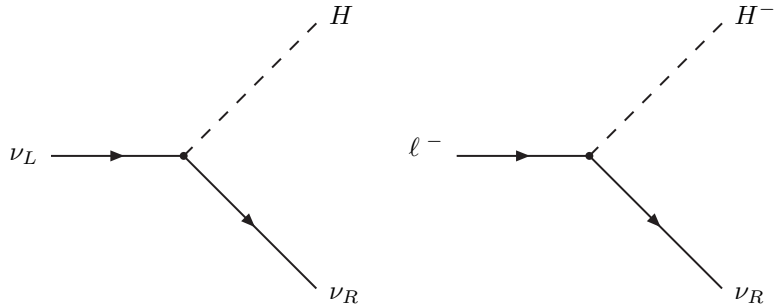

If right-handed bulk fermions are responsible for the small mass of a
neutrino, what are the consequences at the LHC? 
Higgs boson production and decay could be measurably altered from SM or
supersymmetry (SUSY) expectations.
Possible couplings are shown in Figs.~\ref{couple1} and~\ref{couple2}. 
We will use the symbol $\nu_R$ to represent the set of right-handed KK
states, including the right-handed zero mode. 
Although right-handed bulk neutrinos would couple to all types of Higgs
bosons, we restrict our discussion to the SM Higgs boson and a charged
doublet of Higgs bosons.

We review the current bounds on the Higgs boson masses.\cite{CSC}
From electroweak fits and a cut-off scale at the 4-dimensional Planck
scale, the SM Higgs boson mass is restricted to $130 < m_H <
180$~GeV\footnote{The neutral Higgs boson will be denoted by $H$ without
the neutral charge indicated.}.  
If new physics appears at a lower mass scale of 1~TeV, the bounds
becomes weaker: $50 < m_H < 800$~GeV.
Since $m_H > 114.4$~GeV from the LEP experiments, we do not consider
SM Higgs boson masses below 100~GeV.   
The region $160 < m_H < 170$~GeV has also recently been exclude by
the Tevatron experiments.\cite{TEVNPH}
For the charged Higgs boson, $m_{H^+} > 79.3$~GeV is allowed from the LEP
experiments and the Tevatron experiments can limit $m_{H^+} > m_t$ for
$\tan\beta < 1$ or $\tan\beta > 40$, where $m_t$ is the top quark mass.   
We will be interested in charged Higgs bosons that can decay to top
quarks and thus limit our considerations to charged Higgs boson masses
above 170~GeV. 

\begin{figure}[htb]
\begin{center}
\begin{picture}(150,140)(-75,-70)
\SetWidth{0.75}
%\Line(-75,-70)( 75,-70)
%\Line( 75,-70)( 75, 70)
%\Line( 75, 70)(-75, 70)
%\Line(-75, 70)(-75,-70)
\DashLine(-50,0)(0,0){5}
\Vertex(0,0){1.5}
\ArrowLine(50,50)(0,0)
\ArrowLine(0,0)(50,-50)
\Text(-55,0)[r]{$H$}
\Text(55,50)[lb]{$\bar{\nu}_L$}
\Text(55,-50)[lt]{$\nu_R$}
\end{picture}
\begin{picture}(150,140)(-75,-70)
\SetWidth{0.75}
%\Line(-75,-70)( 75,-70)
%\Line( 75,-70)( 75, 70)
%\Line( 75, 70)(-75, 70)
%\Line(-75, 70)(-75,-70)
\DashLine(-50,0)(0,0){5}
\Vertex(0,0){1.5}
\ArrowLine(50,50)(0,0)
\ArrowLine(0,0)(50,-50)
\Text(-55,0)[r]{$H^+$}
\Text(55,50)[lb]{$\ell^{\;+}$}
\Text(55,-50)[lt]{$\nu_R$}
\end{picture}
\end{center}
\caption{\label{couple1}Higgs boson decays involving right-handed bulk
neutrinos.}   
\end{figure}
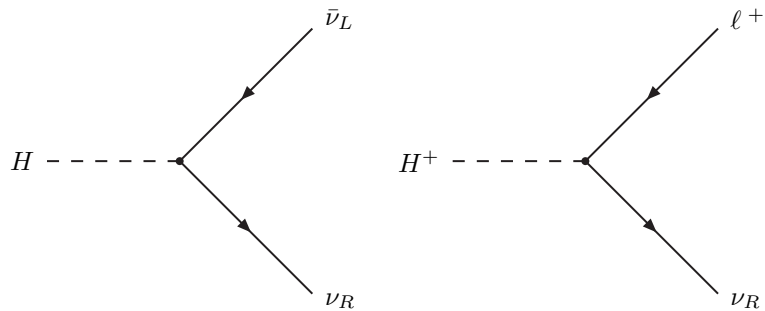

Bulk fermions are not the easiest particles to detect.
The KK states behave as massive, noninteracting, stable particles, and
thus appear as missing energy in the detector. 
The $\nu_L$, $\nu_R^{(0)}$, and KK states all lead to missing energy,
and their detection must be indirect. 

%%%%%%%%%%%%%%%%%%%%%%%%%%%%%%%%%%%%%%%%%%%%%%%%%%%%%%%%%%%%%%%%%%%%%%%%%%%%%%%
\subsection{Tau Decay of a Heavy Charged Higgs Boson}

In this section, we review the possibility of a heavy charged Higgs
boson decaying to a $\tau$ lepton within the model of bulk fermions in
large extra dimensions.   
In typical SUSY models $H^+ \to \tau^+_R \nu_L$ is allowed, while $H^+
\to \tau^+_L \nu_R$ is completely suppressed\footnote{In the following,
the charged Higgs boson is denoted by $H^+$, but the $H^-$ is also
implicitly included.}. 
Models of bulk fermions in large extra dimensions allow for the
possibility of $H^+ \to \tau^+_L \nu_R$, where $\nu_R$ is a singlet
neutrino.   

Many extensions to the SM include a charged Higgs boson.
We will consider the two-Higgs doublet model of type II (2HDM-II).
In this model, each doublet has a unique hypercharge $Y$.
The Higgs doublet with $Y = -1/2$ couples to right-handed up-type quarks
and neutrinos, while the Higgs doublet with $Y = +1/2$ couples to
right-handed down-type quarks and right-handed charged leptons. 
An example of such a model is the Minimal Supersymmetric Standard Model
(MSSM).
In the MSSM there are two vacuum expectation values related by 
$\tan\beta = v_2/v_1$, where $v_2$ is the VEV of the $-1/2$
doublet and $v_1$ is the VEV of the $+1/2$ doublet. 
The VEV $v = 246$~GeV is given by $v/\sqrt{2} = \sqrt{v_1^2 +
v_2^2}$. 

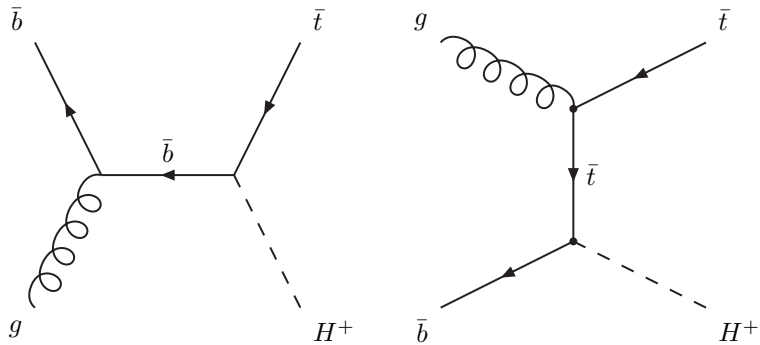
\begin{figure}[htb]
\begin{center}
\begin{picture}(150,140)(-75,-70)
\SetWidth{0.75}
%\Line(-75,-70)( 75,-70)
%\Line( 75,-70)( 75, 70)
%\Line( 75, 70)(-75, 70)
%\Line(-75, 70)(-75,-70)
%\Vertex(0,-25){1.5}
%\Vertex(0,25){1.5}
\Gluon(-50,-50)(-25,0){5}{4}
\ArrowLine(-25,0)(-50,50)
\ArrowLine(25,0)(-25,0)
\ArrowLine(50,50)(25,0)
\DashLine(25,0)(50,-50){5}
\Text(-55,-55)[rt]{$g$}
\Text(-55,55)[rb]{$\bar{b}$}
\Text(55,-55)[lt]{$H^+$}
\Text(55,55)[lb]{$\bar{t}$}
\Text(0,5)[b]{$\bar{b}$}
\end{picture}
\begin{picture}(150,140)(-75,-70)
\SetWidth{0.75}
%\Line(-75,-70)( 75,-70)
%\Line( 75,-70)( 75, 70)
%\Line( 75, 70)(-75, 70)
%\Line(-75, 70)(-75,-70)
\Vertex(0,-25){1.5}
\Vertex(0,25){1.5}
\ArrowLine(0,-25)(-50,-50)
\Gluon(-50,50)(0,25){5}{4}
\ArrowLine(0,25)(0,-25)
\ArrowLine(50,50)(0,25)
\DashLine(0,-25)(50,-50){5}
\Text(-55,-55)[rt]{$\bar{b}$}
\Text(-55,55)[rb]{$g$}
\Text(55,-55)[lt]{$H^+$}
\Text(55,55)[lb]{$\bar{t}$}
\Text(5,0)[l]{$\bar{t}$}
\end{picture}
\end{center}
\caption{\label{charge}Leading-order diagrams for heavy charged Higgs
boson production in association with a top quark.}   
\end{figure}

In the framework of large extra dimensions with bulk fermions, there
is no need to postulate additional Higgs bosons beyond the two
doublets. 
Thus the charged Higgs boson is produced identically as in the 2HDM-II.
For a light charged Higgs boson, the dominant production mechanism is
through the decay of the top quark $t \to H^+ b$.
To focus the discussion, we consider only a heavy charged Higgs boson
with mass larger than the top quark mass.
% $m_{H^+} \gtsimeq m_t$.
We will consider the $2\to 2$ production process $g \bar{b} \to \bar{t}
H^+$, as shown in Fig.~\ref{charge}.  
Charged Higgs boson production in association with a top quark is the
dominant process 
and thus we ignore the processes $b\bar{b} \to H^+W^-$, $b\bar{b} \to
H^+ H^-$, $gg \to H^+ H^-$, and $q\bar{q} \to H^+ H^-$, where $q$ is a
light quark.  

\begin{figure}[htb]
\begin{center}
\includegraphics[width=\columnwidth]{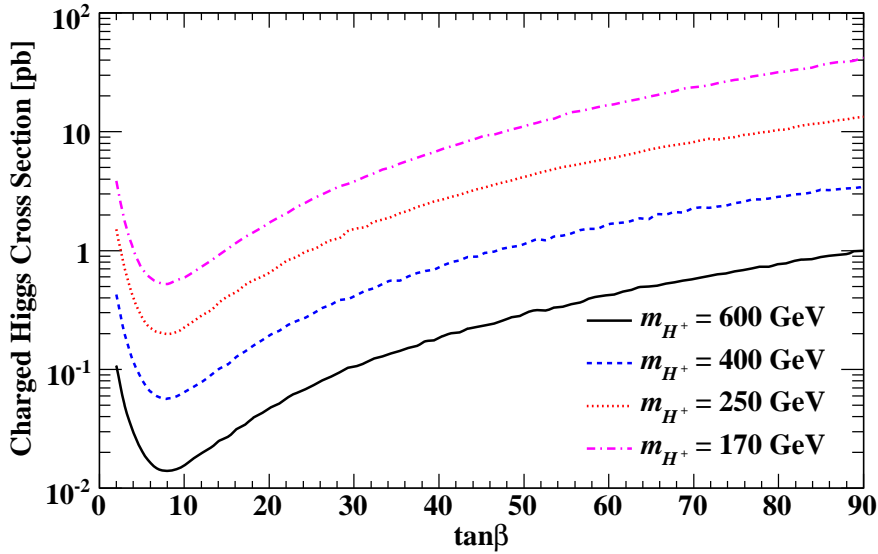}
\caption{\label{xsecch}Leading-order proton-proton cross section at
centre of mass energy of 14~TeV for charged Higgs boson production with
mass above the top quark mass in the MSSM. 
The effects of SUSY particles has been ignored.} 
\end{center}
\end{figure}

Figure~\ref{xsecch} shows the leading-order cross section for charged
Higgs boson production for masses above the top quark mass.
The Yukawa and SUSY electroweak corrections have been calculated along
with the one-loop SUSY corrections.
The complete next-to-leading order (NLO) QCD corrections have been
calculated and are significant. 
The NLO SUSY-QCD corrections are small in comparison.
The next-to-next-to-leading order (NNLO) corrections have been
calculated near threshold.\cite{Plehn03}$^-$\cite{Dittmaier09} 

The $2\to 3$ process $gg \to \bar{t} b H^+ $ is also possible.
This and $g \bar{b} \to \bar{t} H^+$
overlap when summing the theoretical contributions.
Thus one must be careful to avoid double counting when tagging $b$ jets.
Alwall and Rathsman\cite{Alwall04} have addressed this problem and
provided code\cite{Alwall05} to handle the proper matching in the Monte
Carlo generation of events. 
Double counting has been avoided in Fig.~\ref{xsecch} by using
Ref.~\refcite{Alwall05}.

\begin{figure}[htb]
\begin{center}
\includegraphics[width=\columnwidth]{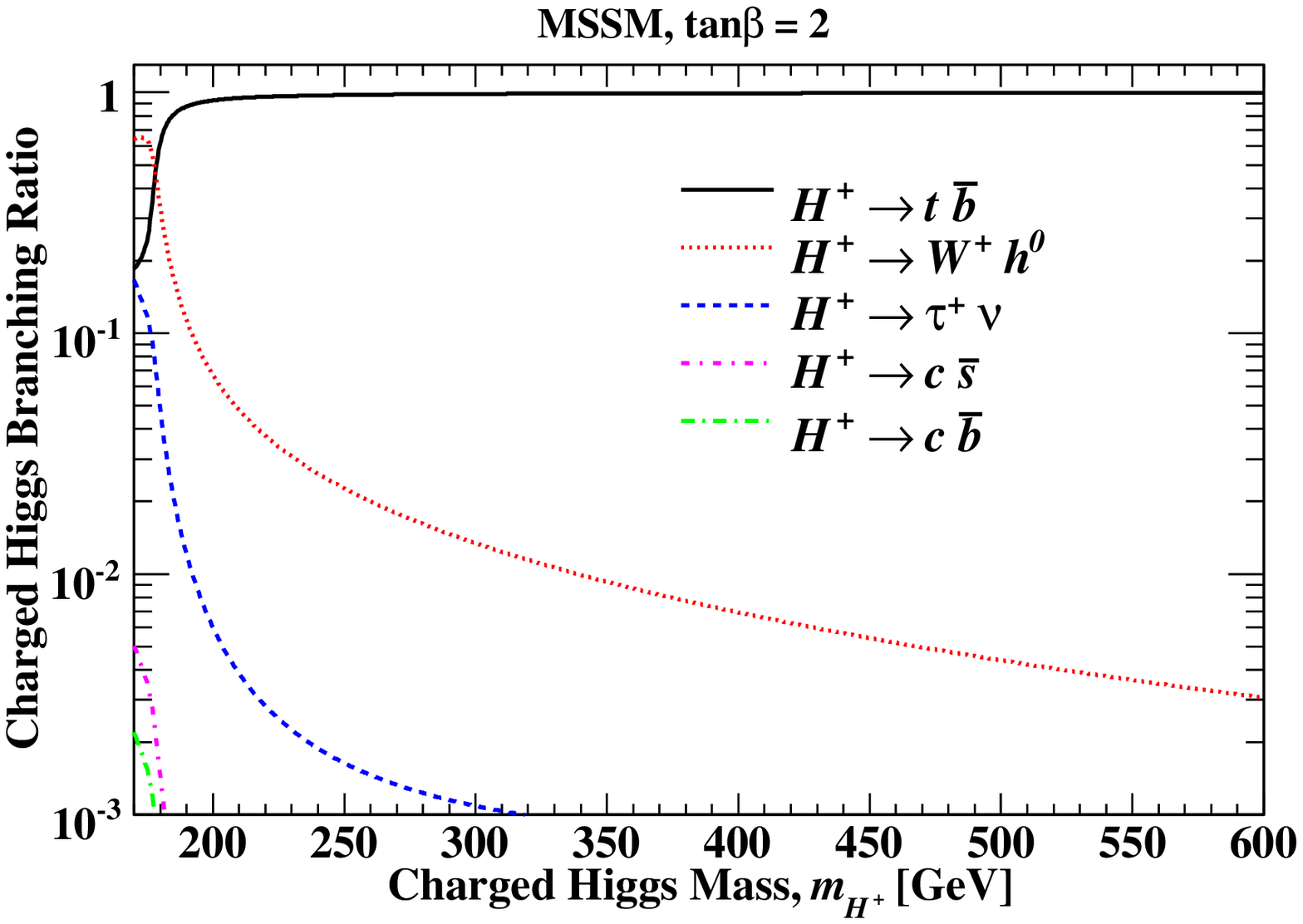}
\includegraphics[width=\columnwidth]{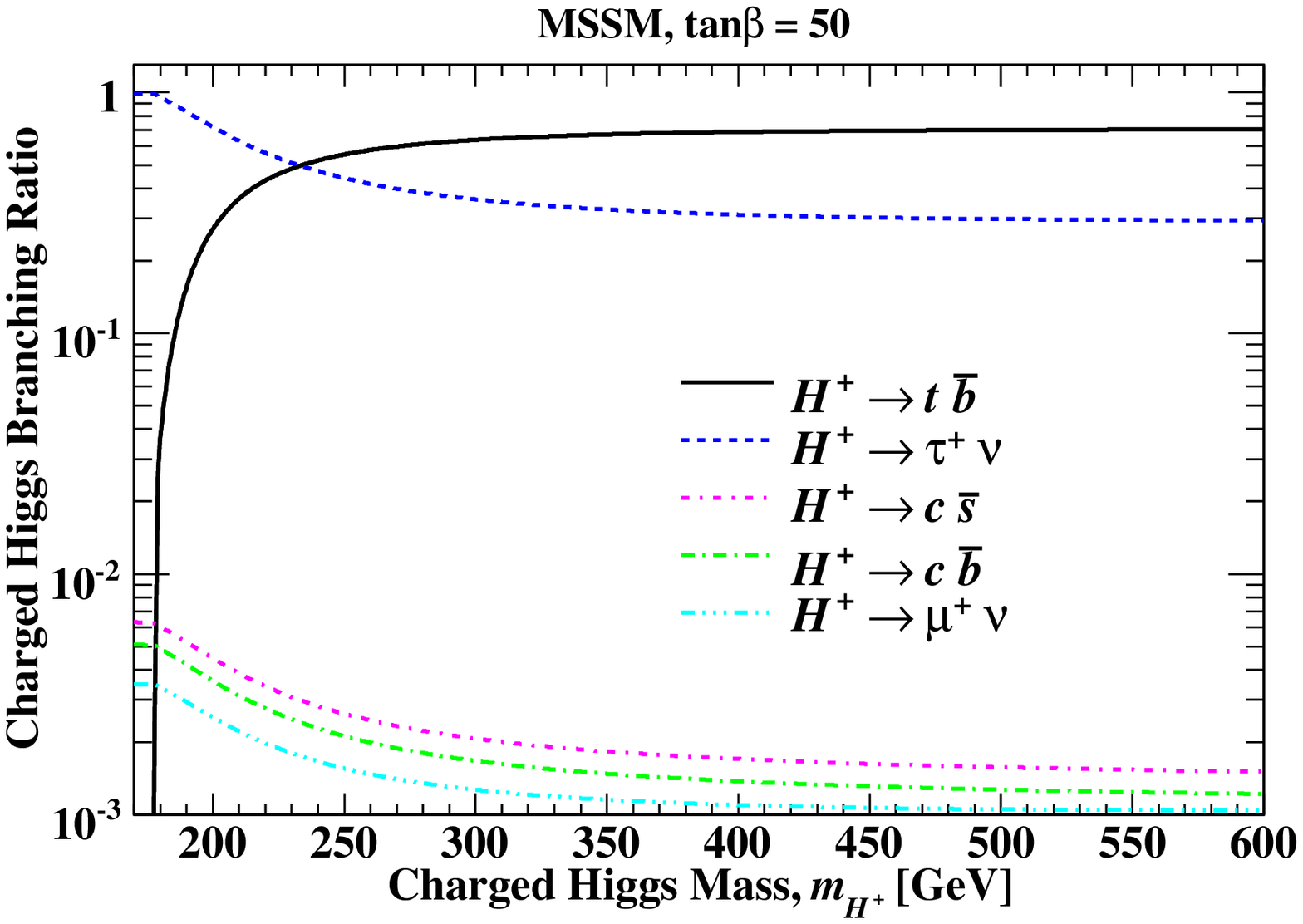}
\caption{\label{br}Branching ratios for charged Higgs boson decays as a
function of charge Higgs boson mass in the MSSM for $\tan\beta = 2$
(top) and $\tan\beta = 50$ (bottom).
Only decays with branching ratios greater than 0.001 in the charged
Higgs boson mass range $170-600$~GeV are shown.}    
\end{center}
\end{figure}

Branching ratios versus the mass of the charged Higgs boson for two
different values of $\tan\beta$ in the MSSM are shown in Fig.~\ref{br}. 
The program HDECAY with default parameters has been used.\cite{HDECAY}
Above the top quark mass threshold, the $H^+ \to t \bar{b}$ decay mode
dominates.
This mode suffers from a large irreducible background and a large
combinatorial background.
A fraction of heavy charged Higgs bosons are allowed to decay into
other modes, depending on the SUSY parameters. 
Particularly, the decay modes $H^+ \to \tau^+ \nu$ and $H^+ \to W^+ h^0$
can be important.  
The discovery potential is dominated by $H^+\to \tau^+ \nu$, which
despite its significantly smaller branching ratio for low $\tan\beta$,
allows more efficient background suppression. 
For high charged Higgs boson masses, decays to SUSY particles may be
kinematically allowed, but we do not consider this possibility. 
This paper assumes the mass scale of the SUSY partners are above
$m_{H^+}$ so that the decays of $H^+$ to SUSY particles are forbidden.  

In the following, we focus on the process 

\begin{equation}
g \bar{b} \to \bar{t} (\to \bar{b} W^-)\ H^+ (\to \tau^+ \nu)\,.
\end{equation}

\noindent
The charged Higgs boson decay to a $\tau$ lepton is the strongest one
affected by the model of bulk fermions in large extra
dimensions.\cite{Agashe00}   

Although both the $W$ boson and charged Higgs boson can decay to $\tau$
leptons, there are important differences in their decays.  
Unlike the $W$ boson that couples universally to all three leptons,
the charged Higgs boson couples preferentially to the heaviest lepton
($\tau$ lepton).   
Since the charged Higgs boson is a scalar particle, the helicities of
the final state particles must be different.
This is opposite to the decay of the $W$ boson, which is a vector
particle. 
Thus, in the MSSM, $H^+ \to \tau^+_R \nu_L$ is allowed but $H^+ \to
\tau^+_L \nu_R$ is forbidden (at tree level) since SM neutrinos are
left-handed.  

For the charged Higgs the corresponding interaction term in the action
corresponding to Eq.~\ref{eq17} is 

\begin{equation}
S_\mathrm{int} = -\frac{\sqrt{2}}{v} \int d^4x \left[ m_D \cot\beta
  \bar{\tau}_L H^+ \psi_R + m_\tau \tan\beta \bar{\tau}_R H^+ \nu_L
  \right] + H.c.\, . 
\end{equation}

\noindent
The corresponding couplings are given in \ref{appendB}.
Thus, in models of large extra dimensions with bulk fermions, both
$\tau$ lepton polarizations are allowed: the usual MSSM mode $\tau^+_R 
\nu$\footnote{For the remainder of this subsection, we drop the $L$
subscript on $\nu_L$ and consider $\nu$ to be either the usual
neutrino of the SM or the left-handed neutrino in models of large
extra dimensions with bulk fermions.} and the new mode $\tau^+_L
\nu_R$, where $\nu_R$ is the right-handed bulk neutrino. 
The usual charged Higgs boson decays to $\tau_R$ leptons through the
Yukawa coupling. 
The SM neutrino $\nu$ is now predominantly a light neutrino plus a small
admixture of KK modes with mass of order $mR/|\hat{n}|$.   
Thus this decay is also effected by the presence of the right-handed
fermion in the extra dimensions.
The decay width $\Gamma(H^+ \to \tau^+_R \nu)$ is modified relative
to the 2HDM-II calculation, $\Gamma_\mathrm{MSSM}$, provided $m_{H^+} <
M_D$ (see \ref{appendB}):\cite{Agashe00}  

\begin{equation}
\Gamma(H^+ \to \tau^+_R \nu) \approx  \frac{1+f}{N^2} 
\Gamma_\mathrm{MSSM}(H^+ \to \tau^+_R \nu)\, , 
\end{equation}

\noindent
where 

\begin{equation}
f \approx \left( \frac{m}{M_D} \right)^2 \left(
\frac{M_\mathrm{Pl}}{M_D} \right)^2 \left( \frac{m_H}{M_D} 
\right)^{\delta-2} x_{\delta-2}\, ,
\end{equation}

\noindent
and 

\begin{equation}
x_{\delta-2} \approx \frac{2\pi^{\delta/2}}{\Gamma(\delta/2)} \left(
\frac{1}{\delta -2} - \frac{2}{\delta} + \frac{1}{\delta +2} \right)\, .
\end{equation}

\noindent
In calculating $f$, the KK states have been summed up to $(m_{H^+}
R)^\delta$, i.e. they are required to be lighter than the charged
Higgs boson mass (see \ref{appendC}).
If  $\delta = 2$, $1/(\delta-2)$ must be replaced by
$\ln(m_{H^+}M_\mathrm{Pl}/M_D^2)$. 
The quantities in the extra factor $(1+f)/N^2$ partially compensate each
other.  
In the parameter space in which $N$ is large, $f$ will also be large, and
when $f$ is small, $N$ is small.  
The $f$ factor varies with charged Higgs boson mass but over the range 
170 to 600~GeV there is little increase in $f$ for low number of
extra dimensions. 
Thus $(1+f)/N^2$ is typically about 0.1 and is always less than unity.

In models of large extra dimensions with bulk fermions, the charged
Higgs boson can also decay to a left-handed $\tau$ lepton and a
right-handed neutrino.   
A calculation of this decay width gives (see \ref{appendB})\cite{Agashe00} 

\begin{equation}
\Gamma(H^+ \to \tau^+_L \nu_R) \approx \frac{m_H}{8\pi} \left(
\frac{m_\mathrm{D}}{v} \right)^2 \cot^2\beta \left( \frac{m_H}{M_D}
\right)^\delta \left( \frac{M_\mathrm{Pl}}{M_D} \right)^2 x_\delta\, ,  
\end{equation}

\noindent
where

\begin{equation}
x_\delta \approx \frac{2\pi^{\delta/2}}{\Gamma(\delta/2)} \left(
\frac{1}{\delta} - \frac{2}{\delta+2} + \frac{1}{\delta+4} \right)\, .  
\end{equation}

\noindent
In the calculation of $\Gamma(H^+ \to \tau^+_L \nu_R)$, the $\tau$
lepton mass has been neglected and the right-handed neutrino KK states
have again been summed up to the threshold for the decays.

In the MSSM at the tree level there are only two parameters:
typically chosen to be the mass of the CP-odd scalar Higgs boson $m_A$
and $\tan\beta$. 
In contrast, the parameter space of a model of large extra dimensions
with bulk fermions depends on $m_\nu$, $m_{H^+}$, $\tan\beta$,
$M_D$, and $\delta$.   
To get a feel for the relative importance of the two helicity modes, the 
ratio of decay widths of left-handed to right-handed $\tau$ leptons is
approximately 

\begin{equation}
x_{LR} \sim \frac{\Gamma(H^+\to \tau^+_L \nu_R)}{\Gamma(H^+\to \tau^+_R
  \nu)} \sim \left( \cot^4\beta \right) \left( \frac{m_\mathrm{D}}{m_\tau}
\right)^2 \left( \frac{M_\mathrm{Pl}}{M_D} \right)^2 \left(
  \frac{m_{H^+}}{M_D} \right)^\delta \, ,  
\end{equation}

\noindent
where for simplicity the normalisation due to mixing $N^2$, and the phase 
space factors $x_\delta$ and $x_{\delta-2}$ have been ignored.
For typical parameter values, $x_{LR} \sim 10^5$.
Part of this high value of $x_{LR}$ is because the decay width to
$\tau_R$ is suppressed by a factor of about 10 relative to the 2HDM-II,
as discussed previously.
The major factor that results in $x_{LR}$ being high is a large
multiplicity factor due to the large number of KK states, despite a
small Yukawa coupling to $\tau_L$ and a single KK neutrino.  
The value of $x_{LR}$ can also be low for high $M_D$ and high
$\tan\beta$. 

\begin{figure}[htb]
\begin{center}
\includegraphics[width=\columnwidth]{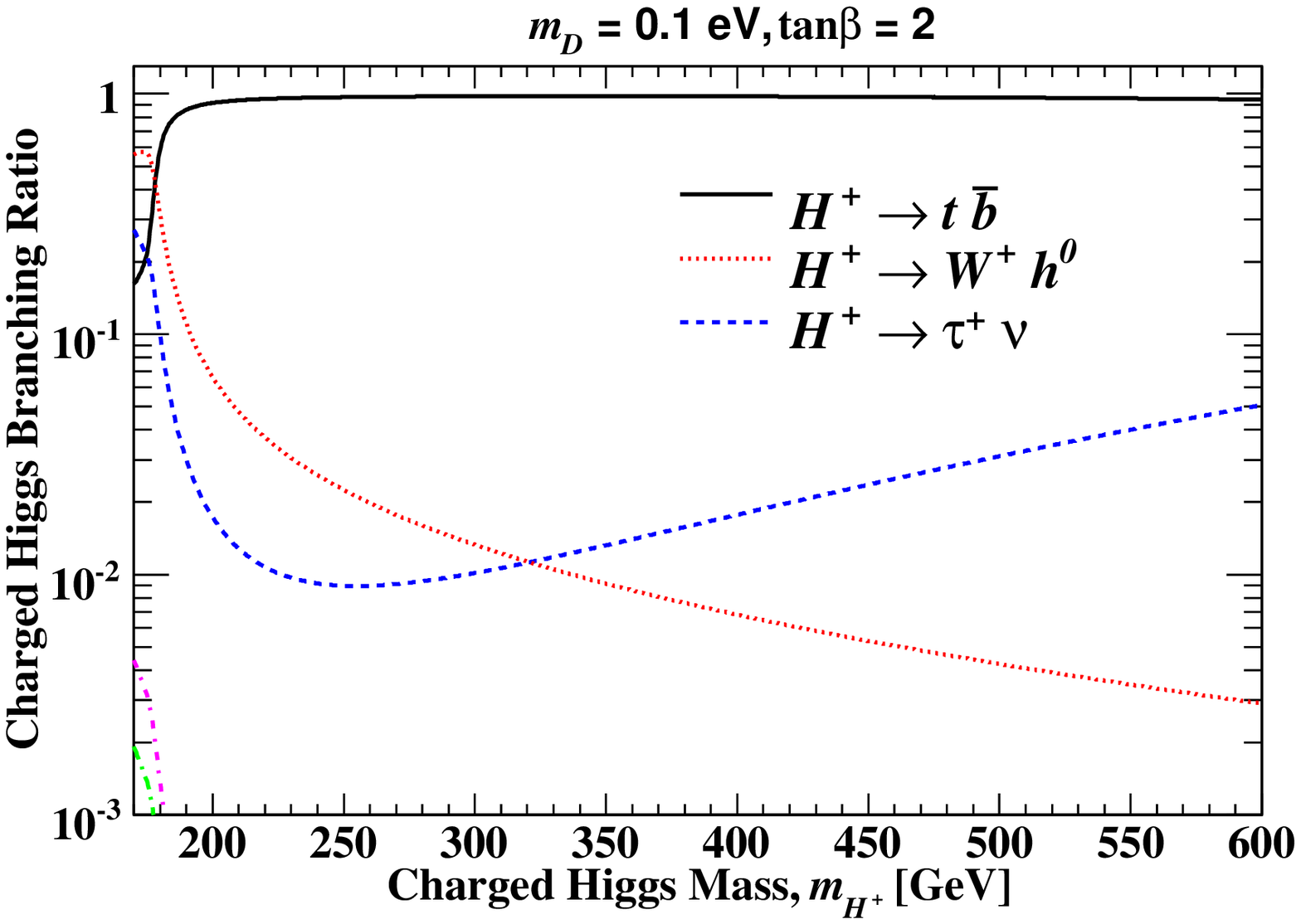}
\includegraphics[width=\columnwidth]{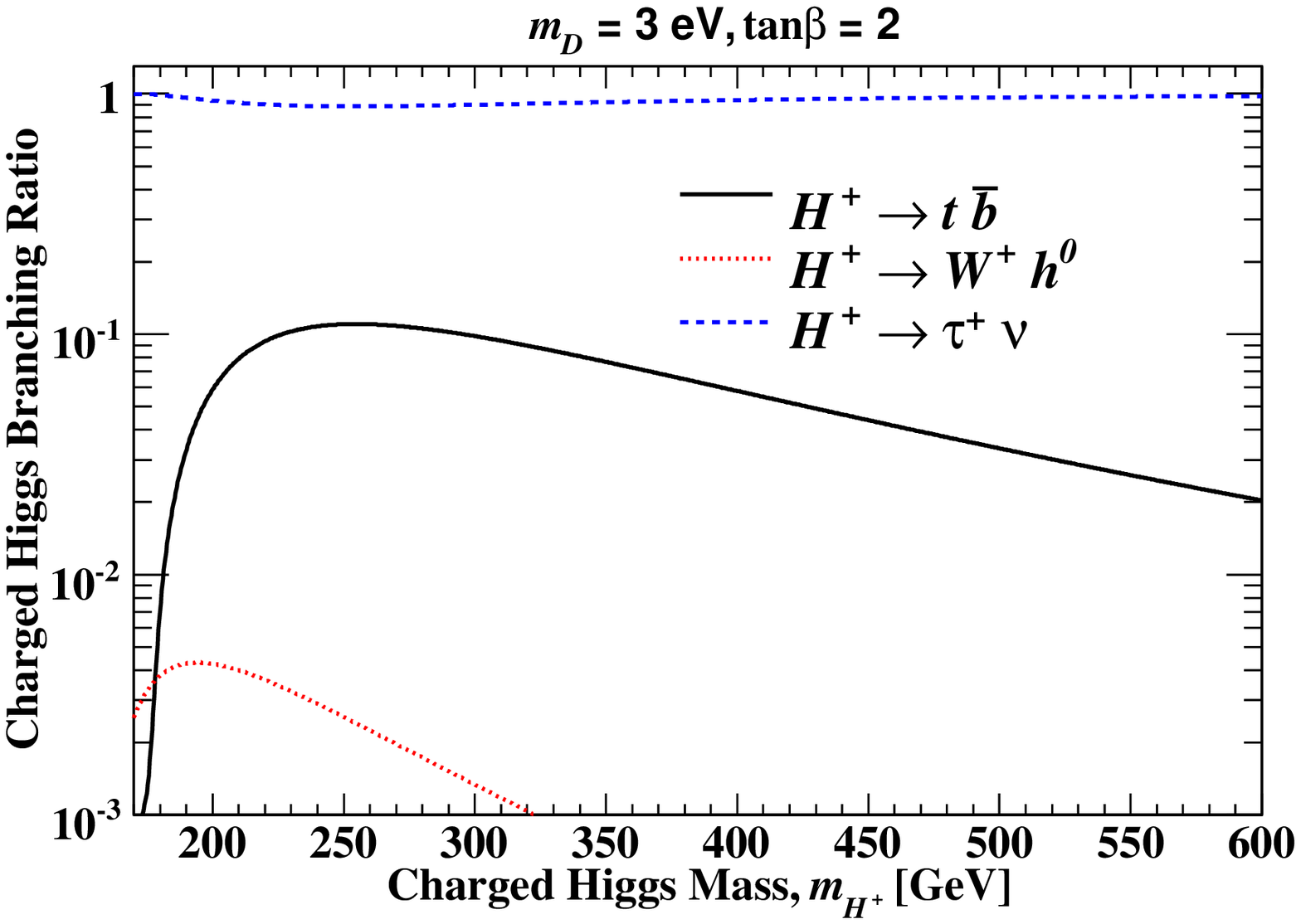}
\caption{\label{tan02}Branching ratios for charged Higgs boson decays as
a function of charge Higgs boson mass for $\tan\beta = 2$, $M_D = 20$~TeV,
$\delta = 3$, and $m_\mathrm{D} = 0.1$~eV (top) and $m_\mathrm{D} =
3$~eV (bottom).  
Only decays with branching ratios greater than 0.001 in the charged
Higgs boson mass range $170-600$~GeV are shown.}    
\end{center}
\end{figure}

\begin{figure}[htb]
\begin{center}
\includegraphics[width=\columnwidth]{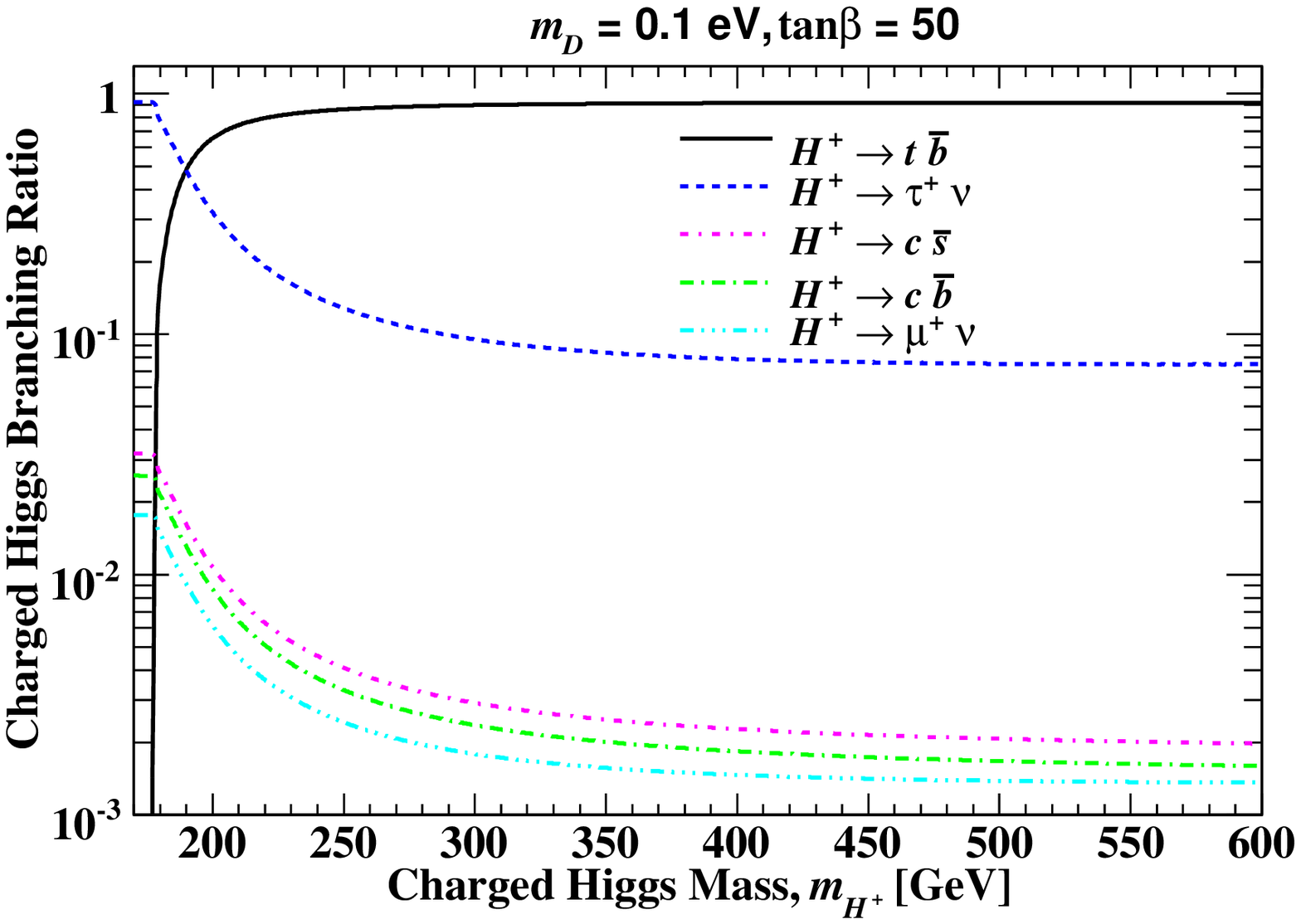}
\includegraphics[width=\columnwidth]{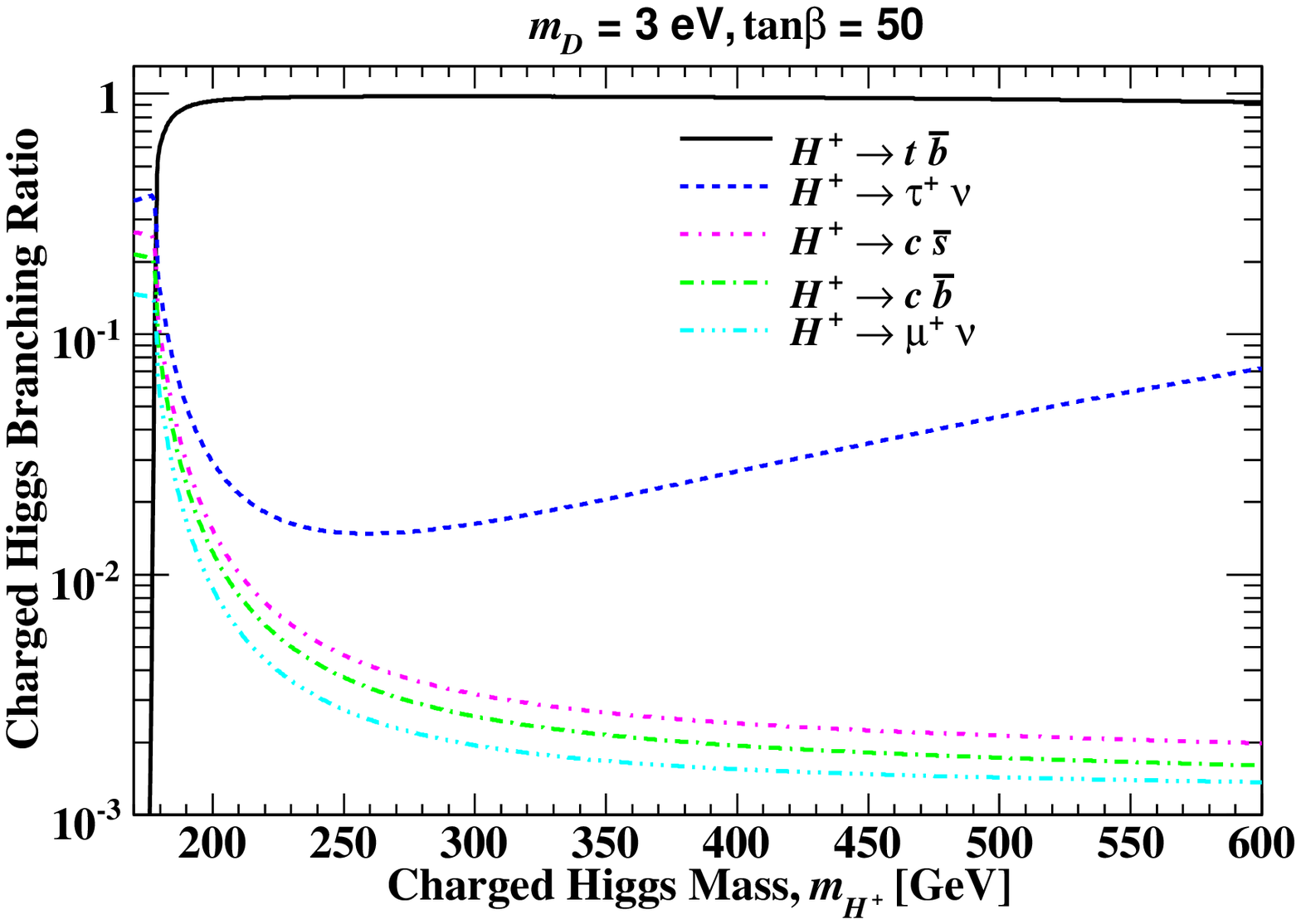}
\caption{\label{tan50}Branching ratios for charged Higgs boson decays as
a function of charge Higgs boson mass for $\tan\beta = 50$, $M_D = 20$~TeV,
$\delta = 3$, and $m_\mathrm{D} = 0.1$~eV (top) and $m_\mathrm{D} =
3$~eV (bottom). 
Only decays with branching ratios greater than 0.001 in the charged
Higgs boson mass range $170-600$~GeV are shown.}    
\end{center}
\end{figure}

Figures~\ref{tan02} and \ref{tan50} show charged Higgs boson branching
ratios for models of large extra dimensions with bulk fermions.
Three extra dimensions was chosen and a value of $M_D = 20$~TeV was used
to ensure the neutrino masses were consistent with the atmospheric
neutrino data. 
Dirac masses of $m_\mathrm{D} = 0.1$~eV (corresponding to $m_\nu^2 =
1.8\times 10^{-3}$~eV$^2$) and $m_\mathrm{D} = 3$~eV (corresponding to
$m_\nu^2 = 2.2\times 10^{-3}$~eV$^2$) have been chosen to lie in the
allowed experimental range.  
For small $m_\mathrm{D}$, we see similar decay branching ratios for
models of large extra dimensions with bulk fermions as in the MSSM.
As $m_\mathrm{D}$ gets larger, $H^+ \to \tau^+ \nu$ becomes dominant for small
$\tan\beta$ values.

In searching for the effects of bulk neutrinos in charged Higgs boson
decay, we need to both extract a signal of charged Higgs boson events
above background events, and show that the resulting signal events agree
with the large extra dimensions scenario.
If one tries to just exploit the $\tau$ lepton polarization difference
to separate bulk fermions in large extra dimension models from the
MSSM, one is probably not able to reduce the background from SM $W$ boson 
decays, as these have the same signature as our extra dimensional model
signal. 
First, we must reduce the SM background to get a clean sample of
charged Higgs boson events. 
After this, an attempt can be made to use the $\tau$ lepton
polarization to determine the charged Higgs boson decay mechanism, and
hence, rule out the MSSM or large extra dimensions model.

One of the backgrounds to our production channel of interest is single
top quark production $g b \to W t$, where the $W$ boson mimics the
charged Higgs boson. 
The cross section times branching ratio for this background is about
48~pb. 
Another background is $t\bar{t}$ production with one $W \to jj$ and the
other $W \to \tau_L \nu$. 
The cross section times branching ratio for this background is about
84~pb.
For the signal, the cross section times branching ratio can be about
0.04~pb to 3.46~pb over the charged Higgs boson mass range of
$200-500$~GeV and $\tan\beta$ from 1.5 to 30.
The background will need to be reduced by a factor of about 40 to 3000. 
The above branching ratio times cross section values have been obtained
from Ref.~\refcite{Assamagan02}.

The differences between signal and background have been examined in 
Refs.~\refcite{Assamagan02,Mohn07}, and more recently, in
Ref.~\refcite{CSC}. 
We largely follow the discussion in Ref.~\refcite{Assamagan02}.
Since we are interested in the $\tau$ lepton decay modes of the 
charged Higgs boson, we require a well reconstructed $\tau$ jet.
One-prong hadronic decays of the $\tau$ lepton are identified and the
reconstructed $\tau$ jet is required to have high transverse momentum 
$p_T^\tau$ to reduce the soft $\tau$ jet backgrounds.
The reason we chose single-prong hadronic $\tau$ lepton decays will be
explained when we discuss measuring the $\tau$ lepton polarization.  

Since the charged Higgs boson is produced in association with a top
quark, we reconstruct single top quark events.
Typically, at least three non-$\tau$ jets with high $p_T$ would be
required. 
One of these jets, and only one, would need to be tagged as a $b$ jet. 
Two of the non-tagged $b$ jets would be required to have an invariant
mass close to the $W$ boson mass.
These two jets would have their energy adjusted to reproduce the $W$
boson mass and the resulting energy-rescaled jets would be combined with
the $b$ jet to reconstruct the three-jet invariant mass.
Events with a three-jet invariant mass close to the top quark mass would
be selected for further analysis.

The difference in mass between the charged Higgs boson and the $W$ boson
can also be exploited.
The $W$ bosons are highly boosted in the laboratory frame of reference
and hence the $\tau$ lepton is approximately collinear with the neutrino
along the direction of the original $W$ boson.  
Thus the distribution of the azimuthal opening angle $\Delta\phi$
between the $\tau$ jet and the missing transverse
energy\footnote{Although we call this quantity missing transverse
energy, we really mean the negative of the vector sum of all the visible
transverse momenta in the event.} $\slashed{E}_T$ 
should be peaked at small values for $W$ boson decays. 
The azimuthal angle $\Delta\phi$ should be larger for $\tau$ leptons
from charged Higgs boson decays.

The difference in mass between the charged Higgs boson and the $W$ boson
can be exploited further.
Because of the neutrino in the final state, only the transverse mass can
be reconstructed:

\begin{equation}
m_T = \sqrt{2p_T^\tau \slashed{E}_T [1 -\cos(\Delta\phi)]}\, .
\end{equation}

\noindent
In background events, the transverse mass has an upper bound at the $W$
boson mass, while in the signal events it is constrained by the charged
Higgs boson mass.  
However, due to finite resolution (particularly on $\slashed{E}_T$),
there is leakage of the background into the signal region.
To optimise the signal to background ratio, a requirement is imposed on
the transverse mass ($m_T > 100$~GeV, for example).
In addition, a cut on $\Delta\phi$ ($\Delta\phi >1.0$, for example) can
be applied. 
Using such a set of requirements as described above and a simple
simulation of the ATLAS detector, Assamagan and
Deandrea\cite{Assamagan02} have shown that a significance (defined by
$S/\sqrt{B}$) greater than five can be obtained with 100~pb$^{-1}$ of
data.  
Using a detailed simulation of the ATLAS detector, the charged Higgs
boson should be detectable in a significant faction of the ($\tan\beta$,
$m_{H^+}$) parameter space with the first 10~fb$^{-1}$ of data.\cite{CSC}
The discovery reach is most likely limited by the signal size
itself.\cite{Assamagan02} 

Observation of a signal in the transverse mass and azimuthal opening
angle distributions could help one to obtain a clean sample of charged
Higgs boson events.
Unfortunately, this would not allow one to determine weather the
scenario is the MSSM or not.  
A further measurement of the polarization asymmetry might provide 
distinctive evidence for models with bulk fermions in large extra
dimensions. 
We can look at the $\tau$ lepton polarization asymmetry which is defined
as  

\begin{equation}
A_\tau = 
\frac{\Gamma(H^+ \to \tau^+_L \nu_R) - 
\Gamma(H^+\to\tau^+_R \nu)}
{\Gamma(H^+\to\tau^+_L \nu_R) +
\Gamma(H^+\to\tau^+_R \nu)}\, . 
\end{equation}

\noindent
In the 2HDM-II, $A_\tau$ is $-1$.
In the model of bulk fermions in large extra dimensions, $A_\tau
\sim -1$ is also allowed but $H^+ \to \tau^+_R \nu$ would have a
different phase space since the neutrino contains an admixture of KK
modes.  
For small values of $m_\nu$ the $\tau$ lepton is right-handed (expect
for small values of $\tan\beta$). 
Left-handed $\tau$ leptons are produced for large $m_\nu$.

To study the $\tau$ lepton polarization, consider the hadronic
single-prong decays  

\begin{equation}
\begin{array}{cclr}
\tau^- & \to & \pi^- \nu                         & \quad\quad (11\%),\\
\tau^- & \to & \rho^- (\to \pi^- \pi^0) \nu      & (25\%),\\
\tau^- & \to & a_1^- (\to \pi^- \pi^0 \pi^0) \nu &  (9\%).\\
\end{array}
\end{equation}

\noindent
These decays best imprint the information of the $\tau$ polarization
onto the decay products.
Experimentally, one does not distinguish between $\pi$ and $K$ mesons,
and $\rho$ and $K^*$ mesons.
Thus a small contribution from the kaon modes is present. 
The above decays correspond to about 90\% of the hadronic one-prong
decays and thus should represent the $\tau$ lepton polarization effects.  
We make the approximation that the decay products of the $\tau$ lepton
merge along the $\tau$ lepton line of flight in the laboratory frame
(collinear approximation).

To take advantage of the direction of the charged $\pi$ meson encoding
the $\tau$ lepton information, the momentum of the $\pi$ meson $p_\pi$
relative to the energy of the $\tau$ jet $E_\tau$,
$x \equiv p_\pi / E_\tau$,
could be examined.
The ratio $x$ is related to the angle that measures the direction of the
charged hadron in the $\tau$ lepton rest frame relative to the $\tau$
lepton line of fight, which defines its polarization
axis.\cite{Raychaudhuri95}  
In the case of MSSM, the $\pi$ mesons coming from charged Higgs boson
decays are peaked at $x=1$.
The distribution is peaked at $x=0$ and $x=1$ for longitudinal $\rho$
and $a_1$ mesons, but is in the middle for transverse $\rho$ and $a_1$
mesons.
In the case of models with bulk fermions or background due to $W\to
\tau_R\nu$ decay, the contributions to the $x$ distribution are
reversed; the $\pi$ mesons coming from the $W$ boson decays are peaked
at $x=0$. 
Since it is anticipated that the discover reach is limited by the signal
size itself, the backgrounds, after the above reduction, are expected to
be very small.
Thus the $x$ distribution could be examined to reveal the model for
charged Higgs boson decays.
The distribution should be peaked at $x=0$ and $x=1$ for the MSSM.
For models of bulk fermions in large extra dimensions, the $x$
distribution would depend on the polarization asymmetry.
In the case of a polarization asymmetry of about one (about 100\%
left-handed $\tau$ leptons) the $x$ distribution would be peaked near
the centre. 

%To take advantage of the direction of the charged $\pi$ meson encoding
%the $\tau$ lepton information, the momentum of the $\pi$ meson $p_\pi$
%relative to the energy of the $\tau$ jet $E_\tau$,
%$x \equiv p_\pi / E_\tau$,
%is required to be large (typically greater than 80\%, say).  
%The ratio $x$ is related to the angle that measures the direction of the
%charged hadron in the $\tau$ lepton rest frame relative to the $\tau$
%lepton line of fight, which defines its polarization
%axis.\cite{Raychaudhuri95}  
%The $\pi$ mesons coming from charged Higgs boson decays are peaked at
%$x=1$, while $\pi$ mesons coming from $W$ boson decays are peaked at
%$x=0$. 
%The distribution is peaked at $x=1$ for longitudinal $\rho$ and $a_1$
%mesons, but is in the middle for transverse $\rho$ and $a_1$ mesons.
%The effect of $\tau$ lepton polarization is reduced by a factor of about
%one half in the $\rho$ meson momentum distribution and practically washed
%out in the case of the $a_1$ meson momentum distribution. 
%Requiring $x$ to be large retains the $\pi$ meson decay modes and only
%about half the $\rho$ and $a_1$ meson contributions.
%For the $\rho$ and $a_1$ mesons, the transverse components have been
%eliminated along with about half of the longitudinal contributions.

In summary, although one could claim a discovery by observing a charged
Higgs boson, it would be insufficient to say what beyond the SM physics
produced it.
Further measurement of the polarization asymmetry might provide
distinctive evidence for the model of bulk fermions in large extra
dimensions. 

%%%%%%%%%%%%%%%%%%%%%%%%%%%%%%%%%%%%%%%%%%%%%%%%%%%%%%%%%%%%%%%%%%%%%%%%%%%%%%%
\subsection{Invisible Decay of a Light Higgs Boson}

Higgs boson invisible decays have been discussed in a number of models
for physics beyond the SM.\cite{Martin99}
To trigger on the invisible decay mode of the Higgs boson, the Higgs
boson must be produced in association with other particles.
In addition, one can not reconstruct the Higgs boson directly but must
probe the signal indirectly through the missing energy distribution.

\begin{figure}[htb]
\begin{center}
\includegraphics[width=\columnwidth]{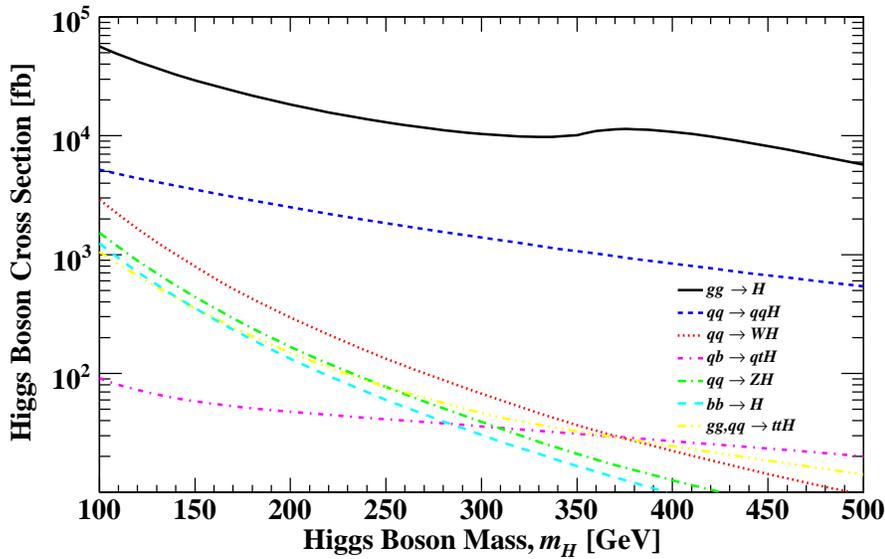}
\caption{\label{xsecsm}Proton-proton cross section at a centre of mass
  energy of 14~TeV for Standard Model Higgs boson production versus
  Higgs boson mass.} 
\end{center}
\end{figure}

For invisible Higgs boson decay in models of bulk neutrinos, no
additional production mechanisms beyond the SM are necessary. 
The cross sections for SM Higgs boson production are shown in
Fig.~\ref{xsecsm}. 
The gluon-gluon fusion channel, via a top quark loop, is dominant.
However, since there is nothing else in the event besides the Higgs
boson, this channel is of little use for triggering on Higgs
boson invisible decays.
The vector-boson vector-boson fusion process ($qq\to qqV^*V^* \to qq H$) has 
recently been shown to be a viable channel for triggering and
searching for Higgs boson invisible decays.\cite{CSC}
Because of its large cross section, it has become a channel of interest.  
The jets are preferentially separated in rapidity and are correlated
in azimuthal angle.
A low-luminosity trigger, which triggers on missing transverse energy
plus a forward jet plus a cental jet is possible. 
At higher luminosity there will be more activity in the rapidity gap
and it is not known how effective the trigger will be.
Further discussion of this channel is beyond the scope of this
phenomenological review.
Other useful channels are the Higgs-strahlung processes
($q\bar{q}^{\,\prime} \to VH$), or associated production with a vector
boson, and associated production with a top quark. 
Associated production with a top quark has not received much attention
and it is anticipated to be less useful then the Higgs-strahlung
processes. 

Associated production with the $W$ boson occurs via $q \bar{q}^{\,\prime}
\to W^* \to W H$, which is followed by $W\to \ell\nu$ and $H \to$
invisible. 
The composite signature to be observed from this decay is 1) no hadronic
activity, 2) missing transverse momentum, and 3) a high-$p_T$ lepton. 
Background due to off-shell $W^*$ production and its leptonic decay
overwhelm the signal channel by a factor of more than
200 even after background rejection cuts.\cite{Godbole03}
We will examine this process in more detail in the next subsection.

A better channel is associated production with the $Z$ boson

\begin{equation} \label{eq46}
q \bar{q}\to Z^* \to Z H\, ,
\end{equation}

\noindent
where $Z\to \ell^+\ell^-$ ($\ell = e$ or $\mu$ only) and $H \to$
invisible. 
The $\tau$ lepton decay mode of the $Z$ boson is usually not considered 
because of uncertainties in $\tau$ lepton identification and the
resulting poor invariant mass resolution from using $\tau$ leptons to
reconstruct $Z$ bosons.  
The signature from the decay is 1) no hadronic activity, 2) missing
transverse momentum, and 3) two high-$p_T$ same flavour charged leptons
with invariant mass close to the mass of the $Z$ boson.
The two high-$p_T$ leptons trigger the event.

There is an irreducible background to the process in Eq.~(\ref{eq46})
from $ZZ$ production, where one $Z$ boson decays leptonically and the
other $Z$ boson decays into neutrinos. 
Since $ZZ$ is produced by $t$-channel processes, it is expected that the
$p_T$ distribution of the $Z$ bosons will be softer than the $p_T$
distribution of the $Z$ bosons from the $ZH$ $s$-channel production
process. 
The next most significant irreducible background is from $WW$ production
with each $W$ decaying leptonically.
This background has a considerably softer transverse momentum
distribution. 
Since both of these backgrounds have softer transverse momentum
distributions than the signal, to may be possible to detect a signal by
requiring high missing transverse energy. 
Other backgrounds arise from $WZ$, $Wj$, and $Z^*\to \tau^+\tau^- \to
\ell^+ \ell^- + \slashed{E}_T$, but they can be suppressed. 

The $Z \to b\bar{b}$ decay channel must also be consider in the
process in Eq.~(\ref{eq46}).
The advantage of this channel is the increased branching fraction of
$Z\to b\bar{b}$ compared to $Z\to \ell^+ \ell^-$.
The disadvantages are the lower efficiency for identifying $b\bar{b}$
final states compared to leptonic final states, the reduced $Z$ boson
invariant mass resolution, and the more difficult background sources.  
These backgrounds include contributions from $ZZ$, $WZ$,
$Zb\bar{b}$, $Wb\bar{b}$, single top quark, and $t\bar{t}$ production.
The significance of the $Z\to b\bar{b}$ channel is not as high as in the
lepton channel, but this channel could be combined with the lepton
channel, or be used to confirm an observed signal.
Another potential background to the process in Eq.~(\ref{eq46}) is $Z\to
\nu\bar{\nu}$ and $H\to b\bar{b}$. 

The SM Higgs boson can decay to $\nu_L$ plus a right-handed bulk
neutrino in models of large extra dimensions. 
The invisible decay width summed over all neutrino flavours is (see
\ref{appendB})\cite{Arkani01,Deshpande03,Martin99} 

\begin{equation}
\sum_{n=0}^{m_HR} \Gamma(H \to \nu_L \bar{\nu}_R^{(n)}) \sim
\sum_{n=0}^{m_HR} \frac{m_H}{4\pi} \left( \frac{m_\mathrm{D}}{v}
\right)^2 \left( 1 - \frac{m^2_{\nu_R^{(n)}}}{m_H^2} \right) d_n \sim
\frac{m_H}{16\pi} \left( \frac{m_\mathrm{D}}{v}\right)^2 (m_H
R)^\delta\, . 
\end{equation}

\noindent
The Higgs boson decay to the final state $\nu_L \bar{\nu}_R$ is proportional
to $(m_\mathrm{D}/v)^2$ which is extremely small. 
However, the multiplicity of KK states below $m_H$ is $(m_H R)^\delta$,
which can be very large.  
It is proportional to the volume $R^\delta$ of the $\delta$-dimensional
space times a momentum-space factor of order $m_H^\delta$.

\begin{figure}[htb]
\begin{center}
\includegraphics[width=\columnwidth]{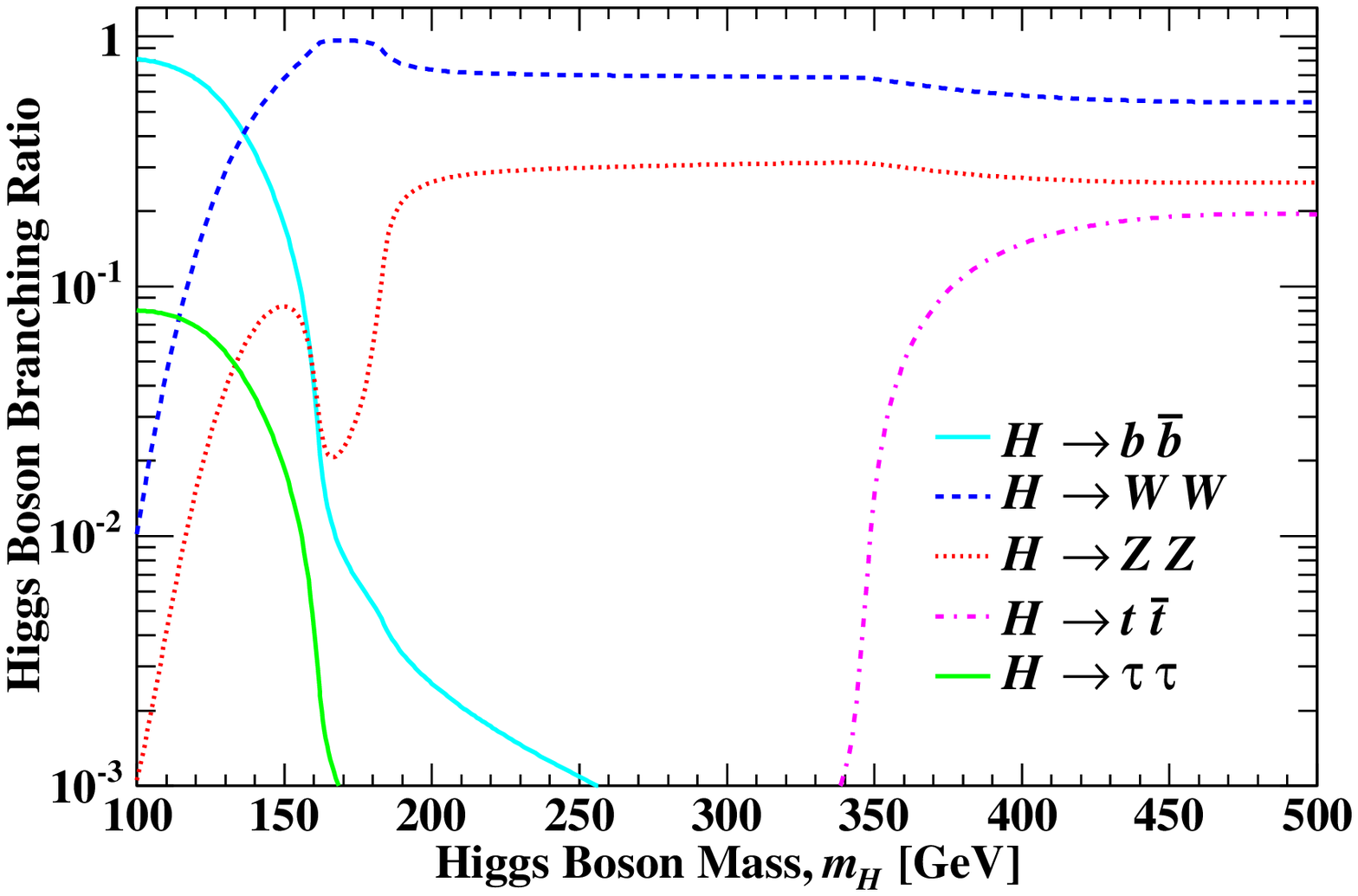}
\includegraphics[width=\columnwidth]{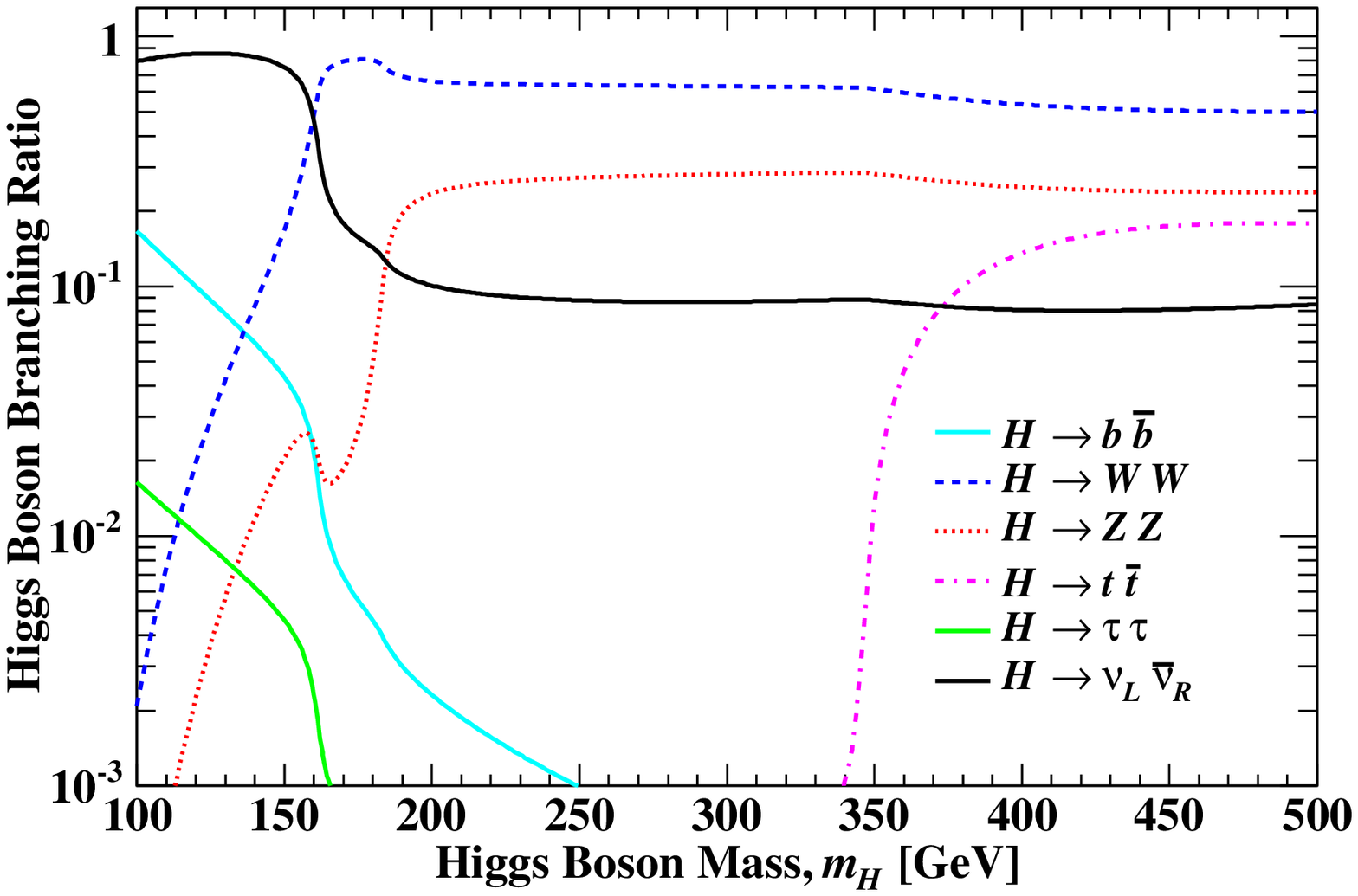}
\caption{\label{invisible}Branching ratios for Standard Model Higgs
boson decays as a function of Higgs boson mass (top) and for a model of
bulk fermions in large extra dimensions with $\delta = 3$, $M_D =
22$~TeV, and $m_\nu^2 = 3\times 10^{-3}$~eV$^2$ (bottom).  
Only decays with branching ratios greater than 0.001 in the Higgs boson
mass range $100-500$~GeV are shown.}     
\end{center}
\end{figure}

Figure~\ref{invisible} shows the Higgs boson branching ratios versus
Higgs boson mass for SM processes and a model with bulk fermions in
large extra dimensions.
Below a Higgs boson mass of 135~GeV the $H\to b\bar{b}$ decay mode
dominates, while above 150~GeV the decay mode $H\to W W^*$ dominates. 

Since $H \to b\bar{b}$ dominates at low Higgs boson masses, it is useful
to consider the ratio

\begin{equation} \label{eq48}
\frac{BR(H\to \nu_L\bar{\nu}_R)}{BR(H\to
b\bar{b})} = \frac{m_\mathrm{D}^2}{3m_b^2} \left( \frac{M_{Pl}}{M_D}
\right)^2 \left( \frac{m_H}{M_D} \right)^\delta\, .
\end{equation}

\begin{figure}[htb]
\begin{center}
\includegraphics[width=\columnwidth]{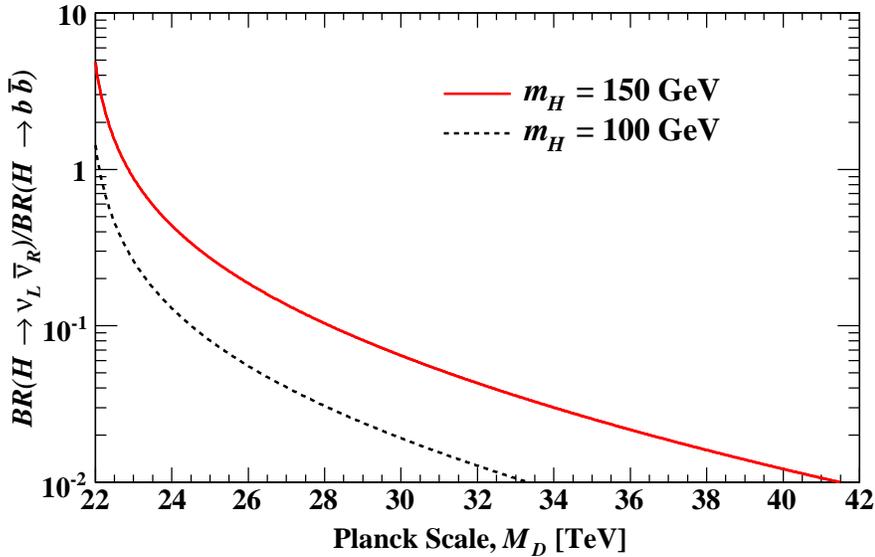}
\caption{\label{bratio}Ratio of Higgs boson invisible decay width to
$b\bar{b}$ decay width versus fundamental Planck scale $M_D$ for two
different Higgs boson masses, and $\delta = 3$ and $m_\nu^2 = 3\times
10^{-3}$~eV$^2$.}   
\end{center}
\end{figure}

\noindent
Figure~\ref{bratio} shows the ratio in Eq.~(\ref{eq48}) versus
fundamental Planck scale. 
The invisible decay mode of the Higgs boson is significant for $\delta =
3$ and low $M_D$.
For $M_D > 30$~TeV, the invisible decay width of the Higgs boson will be
negligibly small compared to $H \to b\bar{b}$. 
One way to reduce $M_D$ is to consider a bulk fermion propagating in 
a subspace $\delta_\nu$ of the full extra dimensions $\delta$ (see
section~\ref{sec3.4}).  
With $\delta_\nu = 5$ and $\delta = 6$, $M_D$ can now be as low as a
TeV.
Only in a very small region of parameter space can the invisible decay
of the Higgs boson be as large as the $H\to b\bar{b}$ decay mode.
The main restriction comes from the perturbative constraint on the
Yukawa coupling $g$. 
Asymmetric dimensions also allow us to reduce $M_D$ while keeping
$g \sim \mathcal{O}(1)$.

\begin{figure}[htb]
\begin{center}
\includegraphics[width=\columnwidth]{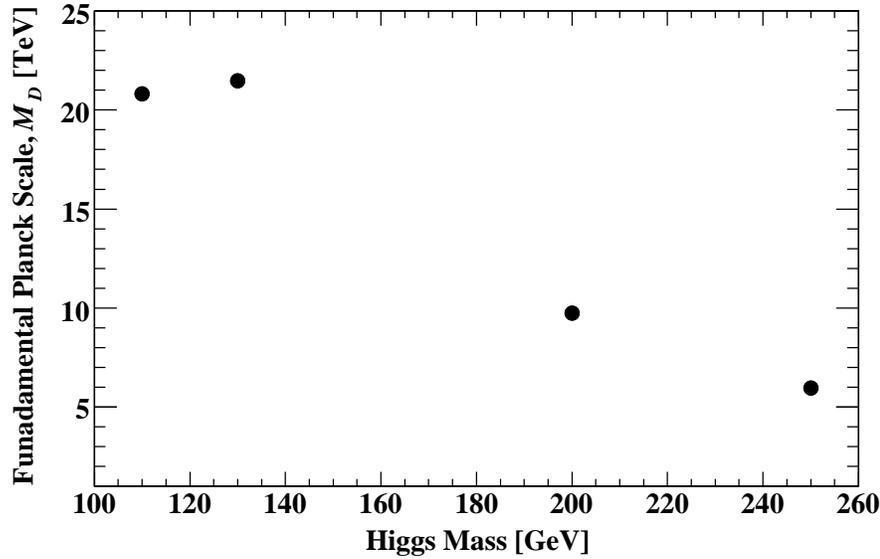}
\caption{\label{invis}Derived 95\% confidence level lower limits on the
fundamental Planck scale $M_D$ versus Higgs boson mass from limits on
an invisible decaying Higgs boson from Ref.~\protect\refcite{CSC},
assuming $\delta = 3$ and $m_\mathrm{D} = 0.1$~eV.} 
\end{center}
\end{figure}

Recently, the ATLAS collaboration has performed a detailed study of its
detector's sensitivity to an invisibly decaying Higgs boson.\cite{CSC}  
Model independent limits were set on the branching ratio time cross
section divided by the SM Higgs boson cross section. 
Assuming the SM Higgs boson production cross section for models of bulk
fermions, limits can be set on the invisibly decay branching ratio as a
function of the neutrino mass, fundamental Planck scale, and the number
of extra dimensions.  
For three extra dimensions and a bulk Dirac fermion mass $m_\mathrm{D} =
0.1$~eV, lower limits can be set on the fundamental Planck scale.
Figure~\ref{invis} show the derived lower limits on the fundamental
Planck scale using the vector boson fusion limits in Ref.~\refcite{CSC}.

In summary, in certain regions of extra-dimensional parameter space, the
branching ratio of Higgs boson decay into invisible modes can be
greater than $BR(H\to b\bar{b})$, but the Yukawa coupling $g$ is large.
For reasonable values of $g$, the invisible decay rate is a tiny
fraction of the $H\to b\bar{b}$ rate.

%%%%%%%%%%%%%%%%%%%%%%%%%%%%%%%%%%%%%%%%%%%%%%%%%%%%%%%%%%%%%%%%%%%%%%%%%%%%%%%
\subsubsection{Associated Production of Bulk Neutrino with Light
Higgs Boson}   

\begin{figure}[htb]
\begin{center}
\begin{picture}(190,140)(-90,-70)
\SetWidth{0.75}
%\Line(-90,-70)(100,-70)
%\Line(100,-70)(100,70)
%\Line(100,70)(-90,70)
%\Line(-90,70)(-90,-70)
\Vertex(-25,0){1.5}
\Vertex(25,0){1.5}
\ArrowLine(-25,0)(-75,-50)
\ArrowLine(-75,50)(-25,0)
\Photon(-25,0)(25,0){5}{5}
\ArrowLine(75,50)(25,0)
\ArrowLine(25,0)(50,-25)
\ArrowLine(50,-25)(75,-50)
\DashLine(50,-25)(75,0){5}
\Text(-80,-55)[rt]{$\bar{q}^{\,\prime}$}
\Text(-80,55)[rb]{$q$}
\Text(80,-55)[lt]{$\nu_R$}
\Text(80,0)[lb]{$H$}
\Text(80,55)[lb]{$\ell^{\;+}$}
\Text(0,10)[b]{$W^*$}
\Text(35,-20)[t]{$\nu_L$}
\end{picture}
\end{center}
\caption{\label{virtual}Associated production of the neutral Higgs boson
with a charged lepton.}   
\end{figure}
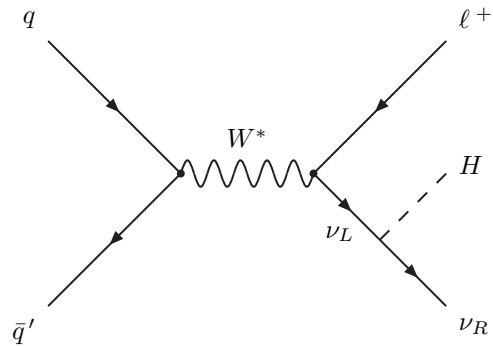

We now discuss in more detail the process of off-shell $W^*$ production
and its leptonic decay.
Consider the process (Fig.~\ref{virtual})

\begin{equation} \label{eq54}
q\bar{q}^{\,\prime} \to W^* \to \ell^+ H \nu_R\, ,
\end{equation}

\noindent
where $\nu_R$ is a KK mass eigenstate.
The process is mediated by $\nu_L$.

In calculating the cross section, we again work in a basis in which the
charged lepton mass matrix is diagonal and $l$ is given by Eq.~(\ref{eq42}). 
Cao, Gopalakrishna, and Yuan\cite{Cao04b} have shown that it is also
beneficial to work in a bases in which $\psi_R^\alpha$ is rotated with
matrix $r$, since this absorbs the unphysical matrix $r$ into the
definition of $\psi_R^\prime$.
It was also found that one should retain $\nu_L$ in the flavour 
basis in order to explicitly keep only the physical matrix $l$.
In this basis, the total production rate for process in Eq.~(\ref{eq54}),
summed over all lepton flavours, is proportional to $\sum_{i,\ell}
|\bar{m}_\mathrm{D}^{i\ell}|^2$, where $\bar{m}_\mathrm{D} \equiv m_\nu^d\,
l^\dag$, and  $m_\nu^d$ is a $3\times 3$ physical diagonalised neutrino
mass matrix. 
Using the neutrino data in section~\ref{sec2}, the matrix
$\bar{m}_\mathrm{D}$ in each mass hierarchy scheme is

\begin{equation}
\bar{m}_\mathrm{D} (\mathrm{normal}) = 
\left(
\begin{array}{ccc}
0 & 0 & 0\\
0.0050 & 0.0053 & -0.0053\\
0 & 0.0354 & 0.0354
\end{array}
\right)\, , 
\end{equation}

\begin{equation}
\bar{m}_\mathrm{D} (\mathrm{inverted}) = 
\left(
\begin{array}{ccc}
0.0414 & -0.0198 & 0.0198\\
0.0280 & 0.0293 & -0.0293\\
0 & 0 & 0
\end{array}
\right)\, ,
\end{equation}

\begin{equation}
\bar{m}_\mathrm{D} (\mathrm{degenerate}) = 
\left(
\begin{array}{ccc}
0.829 & -0.396 & 0.396\\
0.559 & 0.586 & -0.586\\
0 & 0.707 & 0.707
\end{array}
\right)\, ,
\end{equation}

\noindent
and the $\sum_{i,\ell} | \bar{m}_\mathrm{D}^{i\ell}|^2$ values are
(0.0509~eV)$^2$ for the normal, (0.0707~eV)$^2$ for the inverted, and
(1.732~eV)$^2$ for the degenerate mass hierarchy schemes.  

The cross section summed over all lepton flavours $\ell = e, \mu$, and
$\tau$ as a function of $m_\nu$, $m_H$, $1/R$, and $\delta$ has be
calculated in Ref.~\refcite{Cao04b}. 
The cross section is less than a femtobarn for $\delta = 2$, so we will  
focus on $\delta = 3$.  
The cross section decreases slightly with increasing Higgs boson mass.
However, the main dependence is on $1/R$ (or $M_D$) with $\sigma \sim
10^3 - 10^{-3}$~fb for $M_D = 2 - 20$~TeV.
Besides the new physics production mode, there is also the previously 
discussed production mode $q\bar{q} \to W^* \to W(\to \ell\nu) H$, 
which has been show to be indistinguishable from SM background
processes.\cite{Godbole03}  

The Higgs boson can decay to the usual SM decay modes, as well as the
invisible decay mode $H \to \nu_L \bar{\nu}_R$ that we have described
previously.
The SM modes $H\to b\bar{b}$ and $H\to W W^*$ are dominant.
Since the invisible decay mode can only dominate at low Higgs boson
masses $m_H < 160$~GeV, we restrict our attention to this mass region
and will not consider the $H\to WW^*$ decay mode.
It should also be noted that it is difficult to detect the $WW^*$ state.
This is because the leptonic decay branching ratio is suppressed and the
hadronic decay modes are dominated by background.  

We consider the two Higgs boson decay modes, invisible and $b\bar{b}$,
for the new process. 
The decay mode

\begin{equation}
q\bar{q}^{\,\prime} \to W^* \to \ell^+ H(\to b\bar{b})\nu_R
\end{equation}

\noindent
suffers from large background.
However, this channel allows the invariant mass of the Higgs boson to be
reconstructed.
The signature is $\ell b \bar{b}$ plus missing energy, where we will
only consider $\ell = e$ or $\mu$, because of the difficulty in
reconstructing the different decay modes of the $\tau$ jet. 
We will assume the $b$ jets can be tagged by an experiment with an
efficiency of about 50\%, say.
Two intrinsic SM processes dominate the background

\begin{eqnarray}
q \bar{q}^{\,\prime} & \to & W^* \to \bar{b} t [\to b W^+ (\to \ell^+ \nu)
]\, ,\\ 
q \bar{q}^{\,\prime} & \to & W^* \to W^+ (\to \ell^+ \nu) g (\to
b\bar{b})\, . 
\end{eqnarray}

\noindent
Ref.~\refcite{Cao04b} has shown, that even with a perfect detector, for
$\delta = 3$, $M_D = 3.7$~TeV, $m_H = 115$~GeV, and 100~fb$^{-1}$ of 
data,  only eight events above a background of $10^5$ can be obtained.   
Therefore it would be extremely difficult to use the $b\bar{b}$ mode to
directly detect the signal at the LHC.
However, a novel approach employing state-of-the-art jet reconstruction
and decomposition techniques indicates that $H\to b\bar{b}$ is a
promising search channel for a SM Higgs bosons around 120~GeV in
mass.\cite{Butterworth08} 
A recent study by the ATLAS collaboration has confirmed the viability
of the new technique.\cite{Butterworth09}

Now consider the decay mode

\begin{equation}
q \bar{q}^{\,\prime} \to W^* \to \ell^+ H (\to \nu_L\bar{\nu}_R) \nu_R\, .
\end{equation}

\noindent
The signature is $\ell = e, \mu$, or $\tau$ plus missing energy.
For the $\tau$, we will consider only the $\tau^+\to\pi^+\nu$ decay
mode.  
So the resulting signatures are $\pi^+$ plus missing energy or $\ell^+$
plus missing energy, where now $\ell = e$ or $\mu$ only.
The major SM background is the Drell-Yan charged current process 
$q \bar{q}^{\,\prime} \to W^* \to \ell^+ \nu$.

To search for an invisible decay mode, one can apply a set of cuts to
enhance the signal to background and then look at the shape of the
missing transverse energy spectrum.   
The procedure must be applied to the different mass hierarchy schemes
separately since the relative numbers of each lepton type in the final
state differ.
Ref.~\refcite{Cao04b} has shown that the $\ell^+ \slashed{E}_T$ and
$\pi^+ \slashed{E}_T$ channels lead to similar significances.
If $M_D$ is about 5~TeV (for normal or inverted) or about 20~TeV (for 
degenerate) we may expect the significance to be in the range of
$2-5$~$\sigma$.\cite{Cao04b} 
If we strictly impose the constraints on $1/R$, we would expect a poor
significance. 
However, we have pointed out previously that because of the dependence
on the UV cut-off for $\delta=3$, those constraints are uncertain to
some extend. 
Therefore, in the event that the constraint on $M_D$ is relaxed
somewhat, we may expect to see a signal at the LHC.
We may also expect to see a signal if we consider a model with bulk
neutrinos in a subspace of extra dimensions, or asymmetric dimensions.

Neutrino oscillation data do not give the sign of $\Delta
m_\mathrm{atm}^2$ nor the absolute scale of the neutrino masses. 
Measuring the sign of $\Delta m_\mathrm{atm}^2$ would determine
directly whether the normal or the inverted mass hierarchy is realized
in nature. 
Can we use the collider observables to distinguish between the normal,
inverted, or degenerate mass schemes?
We can define a lepton flavour asymmetry.
Let $N(e + \slashed{E}_T)$ and $N(\mu + \slashed{E}_T)$ be the number of
electron and muon signal events, respectively. 
Define

\begin{equation}
A_{\mu e} \equiv \frac{N(\mu + \slashed{E}_T)-N(e +
  \slashed{E}_T)}{N(\mu + \slashed{E}_T) + N(e + \slashed{E}_T)}\, . 
\end{equation}

\begin{figure}[htb]
\begin{center}
\includegraphics[width=\columnwidth]{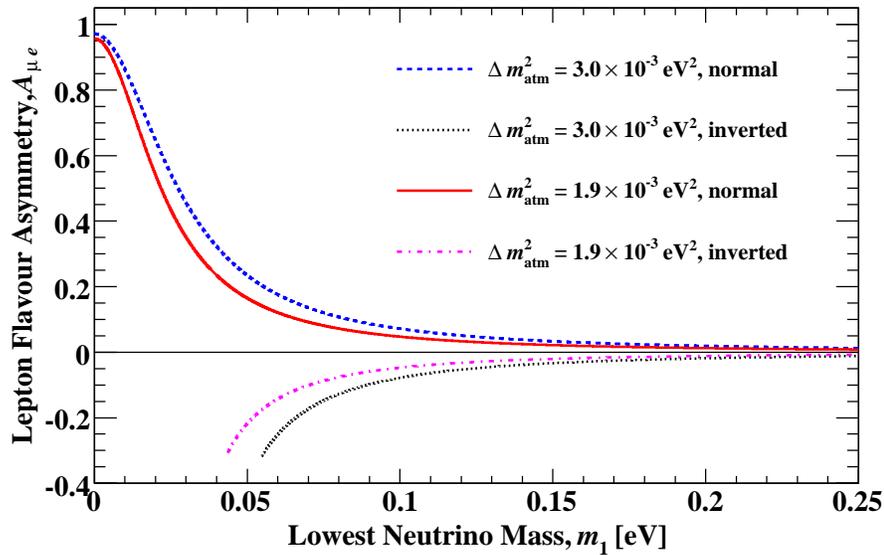}
\caption{\label{hi}Lepton flavour asymmetry $A_{\mu e}$ versus the
lowest neutrino mass eigenstate $m_1$ for the existing bounds on
atmospheric neutrino mixing. $A_{\mu e} > 0$ corresponds to
the normal mass hierarchy, $A_{\mu e} < 0 $ to the inverted mass
hierarchy, and $A_{\mu e} \approx 0$ to the degenerate 
mass hierarchy}   
\end{center}
\end{figure}

\noindent
From this asymmetry one can obtain

\begin{equation}
A_{\mu e} \sim \frac{\pm 0.5 \Delta m_\mathrm{atm}^2}{2m_1^2 \pm 0.5
\Delta m_\mathrm{atm}^2}\, ,
\end{equation}

\noindent
where the upper sign is for the normal mass hierarchy and the lower sign
is for the inverted mass hierarchy.
Thus $A_{\mu e} > 0$ for the normal mass hierarchy and 
$A_{\mu e} < 0$ for the inverted mass hierarchy.
For the inverted mass hierarchy, the smallest value that $m_1$ can take
is 0.05~eV.
Figure~\ref{hi} shows the ideal situation with no experimental affects
included.
For small $m_1$, there is excellent discrimination power to determine
whether the normal or inverted mass hierarchy is realized.
However, as $m_1$ is increased, the number of $e, \mu$, and $\tau$
lepton events becomes approximately equal and the discriminating power 
diminishes. 
We might also be able to determine the absolute scale of the neutrino
mass $m_1$, but only if $m_1^2$ is not too large compared to
$\Delta m_\mathrm{atm}^2$.  
The $\tau-e$ (or $\tau-\mu$) asymmetry does not add any additional
information in probing the neutrino masses.

In summary, a signal of significance $2-5$~$\sigma$ for $\delta=3$ and 
$M_D \sim 5$~TeV (normal or inverted mass hierarchies), or $M_D \sim
20$~TeV (degenerate mass scheme) could be obtained.\cite{Cao04b}
If a positive signal is found, the asymmetry in muon versus electron
events could distinguish between the neutrino mass hierarchies and
determine the absolute neutrino mass scale.
The LHC might be a unique place to determine the hierarchy and mass
scale as it is not available from neutrino oscillation experiments.

%%%%%%%%%%%%%%%%%%%%%%%%%%%%%%%%%%%%%%%%%%%%%%%%%%%%%%%%%%%%%%%%%%%%%%%%%%%%%%%
\section{Summary}

Singlet fermions in the bulk couple to the SM states on the brane as
right-handed neutrinos with small couplings. 
The Yukawa couplings of the bulk fields are suppressed by the large
volume of the extra dimensions.
The interaction between the bulk fields and the brane fields generate
Dirac mass terms between the neutrinos and all the KK modes of the
bulk fields.

The effects of bulk neutrinos could be observed in charged Higgs
boson decays. 
A clean sample of charged Higgs bosons would be needed to identify the
model in which they are realized. 

Observation of Higgs boson invisible decays in models of large extra
dimensions could provide an opportunity to distinguish between the
normal and inverted neutrino mass hierarchies, and to determine the
absolute scale of the neutrino masses. 
Doing this requires measuring the asymmetry of the observed event
numbers in the electrons and muons produced in association with the
Higgs boson. 

Observation of Higgs bosons in models with large extra dimensions will
be difficult.
This is due to the need to keep the dimensionless coupling constant low
enough so that the effective theory is perturbative.  
However, models of asymmetric dimensions or bulk fermions living in
a subspace of the extra dimensions allow a good chance to observe Higgs
bosons at the LHC. 

%%%%%%%%%%%%%%%%%%%%%%%%%%%%%%%%%%%%%%%%%%%%%%%%%%%%%%%%%%%%%%%%%%%%%%%%%%%%%%%
\section*{Acknowledgments}

This work was supported in part by the Natural Sciences and Engineering
Research Council of Canada.

%%%%%%%%%%%%%%%%%%%%%%%%%%%%%%%%%%%%%%%%%%%%%%%%%%%%%%%%%%%%%%%%%%%%%%%%%%%%%%%
\appendix
\section{Mass Mixing Normalisation Factor} \label{appendA}

In this appendix, we perform the sum over KK states in the normalisation
factor of the mass mixing.
From Eq.~(\ref{eq26}),

\begin{equation}
N^2 \approx 1 + \sum_n d_n \left( \frac{mR}{|\hat{n}|} \right)^2 = 1 +
(mR)^2 \sum_n \frac{d_n}{n^2}\, . 
\end{equation}

\noindent
We have dropped the ``hat" notation in the last expression.
For generality, we will not require the number of extra dimensions that
the bulk fermion propagates in $\delta_\nu$ to be the same as the number
of extra dimensions of gravity $\delta$. 
To obtain expressions for the case when the two spaces are the same take
$\delta=\delta_\nu$.

For $\delta_\nu = 1$, the sum can be evaluated directly and the upper
bound on the sum can be taken to infinity. 

\begin{eqnarray}
N^2 & = & 1 + (mR)^2 \frac{\pi^2}{6}\nonumber\\
& = & 1 + \left( \frac{m}{M_D} \right)^2 \left(
\frac{\bar{M}_\mathrm{Pl}}{M_D} \right)^{4/\delta} \frac{\pi^2}{6}\nonumber\\
& = & 1 + \left( \frac{m}{M_D} \right)^2 \left(
\frac{\bar{M}_\mathrm{Pl}}{M_D} \right)^{2\delta_\nu/\delta} \left(
\frac{\bar{M}_\mathrm{Pl}}{M_D} \right)^{2/\delta} \frac{\pi^2}{6}\, ,
\end{eqnarray}

\noindent
where Eq.~(\ref{eq5}) has been used to replace $R$ in terms of $M_D$ and
$\bar{M}_\mathrm{Pl}$ in the second expression.

For $\delta>1$, the infinite sum does not converge so we replace the
sum by an integral over a sphere. 

\begin{equation}
\sum_n d_n \to S_{\delta-1} n^{\delta-1} dn\, ,
\end{equation}

\noindent
where $S_{\delta-1} = 2\pi^{\delta/2} / \Gamma(\delta/2)$.
We have 

\begin{equation}
N^2 \approx 1 + (mR)^2 S_{\delta-1} \int_1^{M_D R} n^{\delta-3} dn\, .
\end{equation}

\noindent
For $\delta_\nu = 2$,

\begin{eqnarray}
N^2 & = & 1 + (mR)^2 2\pi \ln(M_D R)\nonumber\\
& = & 1 + \left( \frac{m}{M_D} \right)^2 \left(
\frac{\bar{M}_\mathrm{Pl}}{M_D} \right)^{4/\delta} 2\pi
\ln\left(\frac{\bar{M}_\mathrm{Pl}}{M_D}\right)^{2/\delta}\nonumber\\
& = & 1 + \left( \frac{m}{M_D} \right)^2 \left(
\frac{\bar{M}_\mathrm{Pl}}{M_D} \right)^{2\delta_\nu/\delta}
\frac{4\pi}{\delta} \ln\left(\frac{\bar{M}_\mathrm{Pl}}{M_D}\right)\, .
\end{eqnarray}

\noindent
For $\delta_\nu > 2$, 

\begin{eqnarray}
N^2 & \approx & 1 + (mR)^2 \frac{2\pi^{\delta_\nu/2}}{\Gamma(\delta_\nu/2)}
\frac{(M_D R)^{\delta_\nu-2}}{\delta_\nu-2}\nonumber\\
& = & 1 + \left( \frac{m}{M_D} \right)^2 (M_D R)^{\delta_\nu}
\frac{2\pi^{\delta_\nu/2}}{\Gamma(\delta_\nu/2)}
\frac{1}{\delta_\nu-2}\nonumber\\ 
& = & 1 + \left( \frac{m}{M_D} \right)^2 \left(
\frac{\bar{M}_\mathrm{Pl}}{M_D} \right)^{2\delta_\nu/\delta}
\frac{2\pi^{\delta_\nu/2}}{\Gamma(\delta_\nu/2)} \frac{1}{\delta_\nu-2}\, .
\end{eqnarray}

\noindent
In the first line of above expression, we have made the approximation that the
lower bound of integration can be taken to be zero.

%%%%%%%%%%%%%%%%%%%%%%%%%%%%%%%%%%%%%%%%%%%%%%%%%%%%%%%%%%%%%%%%%%%%%%%%%%%%%%%
\section{Scalar Boson to Two-Fermion Decay Widths} \label{appendB}

The $1\to 2$ decay width is

\begin{equation}
\Gamma = \frac{p}{8\pi M^2} |\mathcal{M}|^2\, ,
\end{equation}

\noindent
where the magnitude of either outgoing momentum in the boson rest frame
is  

\begin{eqnarray}
%p & = & \frac{1}{2M} \sqrt{ [M^2 - (m_1 + m_2)^2] [M^2 - (m_1 -
%m_2)^2]}\nonumber\\ 
p & = & \frac{M}{2} \sqrt{ \left[ 1 - \left( \frac{m_1 + m_2}{M} \right)^2
\right] \left[ 1 - \left( \frac{m_1 - m_2}{M} \right)^2 \right]}\, .
\end{eqnarray}

%\noindent
%If $m_1 = m_2 \equiv m$,

%\begin{eqnarray}
%p & = & \frac{M}{2} \sqrt{ 1 - 4\left( \frac{m}{M} \right)^2}\, .
%\end{eqnarray}

\noindent
If $m_2 \ll m_1$ or $m_2 \ll M$ ($m_1 \equiv m$),

\begin{eqnarray}
p & \approx & \frac{M}{2}  \left[ 1 - \left( \frac{m}{M} \right)^2
\right]\, .
\end{eqnarray}

\noindent
If $m_1 \ll M$ and $m_2 \ll M$,

\begin{eqnarray}
p & \approx & \frac{M}{2}\, .
\end{eqnarray}

The matrix element consists of a vertex factor times a spinor product 

\begin{equation}
\mathcal{M} = y_\nu \mathcal{M}^{\,\prime}\, .
\end{equation}

The spinor product for a signal polarization state is

\begin{eqnarray}
\mathcal{M}^{\,\prime} & = & \bar{u} v = u^\dag \gamma^0 v = \sqrt{(E_1 +
  m_1)(E_2 - m_2)} + \sqrt{(E_1 - m_1)(E_2 + m_2)}\nonumber\, ,\\
%(\mathcal{M}^{\,\prime})^2 & = & (E_1 + m_1)(E_2 - m_2) + (E_1 - m_1)(E_2 +
%  m_2) + 2 \sqrt{(E_1^2 - m_1^2)(E_2^2 - m_2^2)}\nonumber\\
%& = & 2E_1 E_2 -2 m_1 m_2 + 2p^2\nonumber\\
%& = & 2(M - E_2) E_2 -2 m_1 m_2 + 2p^2\nonumber\\
(\mathcal{M}^{\,\prime})^2 & = & 2 M E_2 - 2 m_2^2 -2 m_1 m_2\, .
\end{eqnarray}

%\noindent
%If $m_1 = m_2 \equiv m$,

%\begin{equation}
%\mathcal{M}^{\,\prime} = 2 p\, .
%\end{equation}

\noindent
If $m_2 \ll M$,

\begin{equation}
\mathcal{M}^{\,\prime} \approx \sqrt{2 M p}\, .
\end{equation}

\noindent
If $m_1 \ll M$ and $m_2 \ll M$,

\begin{eqnarray}
\mathcal{M}^{\,\prime} \approx 2 E = M\, .
\end{eqnarray}

For fermions (not neutrinos), there will be two polarization states and 
hence the matrix element squared should be multiplied by a factor of
two. 
For SM neutrinos this factor of two will be absent.
Including the two polarization states, we have

%\begin{equation}
%\Gamma = \frac{y_\nu^2M}{8\pi} \left[ 1 - 4\left( \frac{m}{M} \right)^2
%\right]^{3/2}\, , 
%\end{equation}

%\noindent
%for $m_1 = m_2 \equiv m$.

\begin{equation}
\Gamma = \frac{y_\nu^2M}{8\pi} \left[ 1 -\left( \frac{m}{M} \right)^2
\right]^2\, ,  
\end{equation}

\noindent
for $m_2 \ll M$.

\begin{equation}
\Gamma = \frac{y_\nu^2M}{8\pi}\, ,
\end{equation}

\noindent
for $m_1 \ll M$ and $m_2 \ll M$.

%%%%%%%%%%%%%%%%%%%%%%%%%%%%%%%%%%%%%%%%%%%%%%%%%%%%%%%%%%%%%%%%%%%%%%%%%%%%%%
\subsection{SM Higgs Boson and MSSM Charged Higgs Boson Decay Widths}
\label{appendB1} 

\noindent
For the SM Higgs boson,

\begin{equation}
y_\nu = -i \frac{gm_f}{2m_W} = -i \frac{m_f}{v}\, .
\end{equation}

\noindent
For $H^+\to t\bar{b}$ in MSSM,

\begin{equation}
y_\nu = \frac{\sqrt{2}}{v} \left( m_t \cot\beta P_R + m_b \tan\beta
P_L \right)\, .
\end{equation}

\noindent
For $H^+\to \tau^+ \nu$ in MSSM,

\begin{equation}
y_\nu = \frac{\sqrt{2}}{v} \left( m_\tau \tan\beta P_L \right)\, .
\end{equation}

Consider $H^+\to\tau^+_R\nu$ in MSSM.
Using 
$y_\nu = \sqrt{2} (m_\tau/v) \tan\beta P_L$, 
$p = M/2$, and
$\mathcal{M}^{\,\prime} = M$ gives

\begin{eqnarray}
%\Gamma^{MSSM} & = & \frac{1}{8\pi M} \left[ 2 \left( \frac{m_\tau}{v}
%\right)^2 \tan^2 \beta \right] M^2 \frac{M}{2}\nonumber\\
\Gamma^{MSSM} & = &  \frac{M}{8\pi} \left( \frac{m_\tau}{v} \right)^2
\tan^2\beta\, . 
\end{eqnarray}

%%%%%%%%%%%%%%%%%%%%%%%%%%%%%%%%%%%%%%%%%%%%%%%%%%%%%%%%%%%%%%%%%%%%%%%%%%%%%%
\subsection{Large Extra Dimensions Decay Widths} \label{appendB2}

Consider $H^+\to\tau^+_R\nu$ in the model of bulk fermions in large
extra dimensions. 
We separate the zero mode from the $\hat{n}$ modes; $\Gamma =
\sum_n \Gamma^{(n)} = \Gamma^{(0)} + \sum_{\hat{n}} \Gamma^{(\hat{n})}$. 
Using 
$y_\nu = \sqrt{2} (m_\tau/v) \tan\beta P_L$, 
$p_0 = M/2$, $p_{(\hat{n})} = (M/2)[1-1/M^2(n/R)^2]$, and
$\mathcal{M}_0^{\,\prime} = M/N$,
$\mathcal{M}_{(\hat{n})}^{\,\prime} =
(m_\mathrm{D}R/n)\sqrt{2Mp_{(\hat{n})}}/N$ gives 

\begin{eqnarray}
\Gamma & = & \frac{\Gamma_{MSSM}}{N^2}\nonumber\\
& &  + \frac{1}{8\pi M} \left[ 2
\left( \frac{m_\tau}{v} \right)^2 \tan^2 \beta \right] \frac{1}{N^2}
\sum_n \left( \frac{m_\mathrm{D}R}{n} \right)^2 2M \frac{M^2}{4} \left[
1 -\frac{1}{M^2} \left( \frac{n}{R} \right)^2 \right]^2 \nonumber\\
& = & \frac{\Gamma_{MSSM}}{N^2} + \frac{M}{8\pi} \left(
\frac{m_\tau}{v} \right)^2 \tan^2\beta \frac{1}{N^2} \left(
\frac{m_\mathrm{D}}{M} \right)^2 \sum_n M^2 \left( \frac{R}{n} \right)^2
\left[1 -\frac{1}{M^2} \left( \frac{n}{R} \right)^2 \right]^2\nonumber\\  
& = & \frac{\Gamma_{MSSM}}{N^2} \left[ 1 + \left( \frac{m_\mathrm{D}}{M}
\right)^2 (MR)^\delta x_{\delta-2}\right]\nonumber\\  
& = & \frac{\Gamma_{MSSM}}{N^2} \left[ 1 + \left( \frac{m_\mathrm{D}}{M_D}
\right)^2 \left( \frac{M_\mathrm{Pl}}{M_D} \right)^2 \left(
\frac{M}{M_D} \right)^{\delta-2} x_{\delta-2}\right]\, .   
\end{eqnarray}

\noindent
See \ref{appendC} for a calculation of $x_{\delta-2}$.

Consider $H^+\to\tau^+_L\nu_R$.
Using 
$y_\nu = \sqrt{2} (m_\mathrm{D}/v) \cot\beta P_R$, 
$p = (M/2)[1 -(1/M^2)(n/R)^2]$, and
$\mathcal{M}^{\,\prime} = \sqrt{2Mp}$ gives

\begin{eqnarray}
\Gamma & = & \frac{1}{8\pi M} \left[ 2 \left( \frac{m_\mathrm{D}}{v}
\right)^2 \cot^2 \beta \right] 2 M \frac{M^2}{4} \sum_n \left[ 1
-\frac{1}{M^2} \left( \frac{n}{R} \right)^2 \right]^2\nonumber\\
& = &  \frac{M}{8\pi} \left( \frac{m_\mathrm{D}}{v} \right)^2 \cot^2\beta
(MR)^\delta x_\delta\, .
\end{eqnarray}

\noindent
See \ref{appendC} for a calculation of $x_\delta$.

%%%%%%%%%%%%%%%%%%%%%%%%%%%%%%%%%%%%%%%%%%%%%%%%%%%%%%%%%%%%%%%%%%%%%%%%%%%%%%%
\section{Kaluza Klein Phase Space Sums} \label{appendC}

For bulk fermions, the masses of the KK states are not negligible,
nor are the masses of the final two particles equal. 
Thus $[1 - (m/M)^2]^2$ with $m = n/R$ summed over the number of KK
states is the quantity of interest in the decay width. 

\begin{eqnarray}
x_\delta & \equiv & \sum_n \left[ 1 - \frac{1}{M^2} \left(
\frac{n}{R}\right)^2 \right]^2\nonumber\\
& = & \sum_n \left[ 1 - \frac{2}{(MR)^2} n^2 + \frac{1}{(MR)^4} n^4
\right]\nonumber\\ 
& = & \frac{2\pi^{\delta/2}}{\Gamma(\delta/2)} \int_0^{MR} \left[
n^{\delta-1} - \frac{2}{(MR)^2} n^{\delta+1} + \frac{1}{(MR)^4}
n^{\delta+3} \right] dn\nonumber\\
& = & \frac{2\pi^{\delta/2}}{\Gamma(\delta/2)} \left[
\frac{n^\delta}{\delta} -
\frac{2n^{\delta+2}}{(MR)^2(\delta+2)} +
\frac{n^{\delta+4}}{(MR)^4(\delta+4)} \right]_0^{MR}\nonumber \\   
& = & (MR)^\delta \frac{2\pi^{\delta/2}}{\Gamma(\delta/2)} \left(
  \frac{1}{\delta} - \frac{2}{\delta+2} + \frac{1}{\delta+4} \right)\, .
\end{eqnarray}

\noindent
Also of interest is

\begin{eqnarray}
x_{\delta-2} & \equiv & \sum_n M^2 \left( \frac{R}{n} \right)^2 \left[ 1 -
  \frac{1}{M^2} \left( \frac{n}{R}\right)^2 \right]^2\nonumber\\
& = & \sum_n   \left[ (MR) \frac{1}{n} -
\frac{1}{(MR)} n\right]^2\nonumber\\ 
& = & \sum_n \left[ (MR)^2 \frac{1}{n^2} - 2 + \frac{1}{(MR)^2} n^2
\right]\nonumber\\ 
& = & \frac{2\pi^{\delta/2}}{\Gamma(\delta/2)} \int_0^{MR} \left[ (MR)^2
n^{\delta-3} - 2 n^{\delta-1} + \frac{1}{(MR)^2} n^{\delta+1} \right]
dn\nonumber\\  
& = & \frac{2\pi^{\delta/2}}{\Gamma(\delta/2)} \left[
(MR)^2 \frac{(MR)^{\delta-2}}{\delta-2} - 2 \frac{(MR)^\delta}{\delta} +
\frac{1}{(MR)^2} \frac{(MR)^{\delta+2}}{\delta+2} \right]\nonumber \\   
& = & (MR)^\delta \frac{2\pi^{\delta/2}}{\Gamma(\delta/2)} \left(
\frac{1}{\delta-2} - \frac{2}{\delta} + \frac{1}{\delta+2} \right)\, .
\end{eqnarray}

%%%%%%%%%%%%%%%%%%%%%%%%%%%%%%%%%%%%%%%%%%%%%%%%%%%%%%%%%%%%%%%%%%%%%%%%%%%%%%%

\end{document}